\documentclass[a4paper,11pt]{article}

\usepackage{geometry}										% [showframe]
\usepackage[english]{babel}									% language
\usepackage[utf8]{inputenc}									% input text encoding
\usepackage[T1]{fontenc}									% output encoding
\usepackage{graphicx}
\usepackage{float}

\usepackage{lmodern}										% Latin Modern

\usepackage{bm}
\usepackage{amsmath, amssymb}
\usepackage{xcolor, soul}
\usepackage{booktabs}

\makeatletter
\let\oldabs\abs
\def\abs{\@ifstar{\oldabs}{\oldabs*}}
\let\oldnorm\norm
\def\norm{\@ifstar{\oldnorm}{\oldnorm*}}
\makeatother

\DeclareMathOperator*{\argmax}{arg\,max}

\newcommand{\xb}{\bm{x}}

\newcommand{\mub}{\boldsymbol{\mu}}
\newcommand{\Xb}{\bm{X}}
\newcommand{\thb}{\boldsymbol{\theta}}
\newcommand*{\tran}{^{\mkern-1.5mu\mathsf{T}}}
\newcommand{\zetab}{\boldsymbol{\zeta}}

\usepackage{authblk}										% [auth-sc,affil-it], to be loaded after titling
\title{Comparison of Probabilistic Structural Reliability Methods for Ultimate Limit State Assessment of Wind Turbines}
\author[1]{Hong Wang}
\author[1]{Odin Gramstad}
\author[2]{Styfen Schär \thanks{styfen.schaer@ibk.baug.ethz.ch}}
\author[2]{Stefano Marelli\thanks{marelli@ibk.baug.ethz.ch}}
\author[3]{Erik Vanem\thanks{Erik.Vanem@dnv.com}}

\affil[1]{DNV Energy, H{\o}vik, Norway}
\affil[2]{Chair of Risk, Safety and Uncertainty Quantification, ETH Z\"{u}rich, Switzerland}
\affil[3]{DNV Group Research and Development, H{\o}vik, Norway}

\date{\today}

\usepackage{microtype}										% typographical refinements

\usepackage{indentfirst}									% indents first paragraph in new section

\usepackage{caption}
\usepackage{subcaption}

\usepackage{natbib}
\usepackage{hyperref}
\hypersetup{
	pdftitle = {Comparison of Probabilistic Structural Reliability Methods for Ultimate Limit State Assessment of Wind Turbines},
	pdfauthor = {Hong Wang, Odin Gramstad, Styfen Sch\"ar, Stefano Marelli, Erik Vanem},
	pdfsubject = {},
	pdfkeywords = {Structural reliability, Probabilistic design, Offshore wind, Environmental contours, Sequential sampling, Surrogate models, Renewable energy},
	colorlinks = false,                                                                                                                                                                          % false: boxed links; true: colored links
}

\usepackage[capitalize]{cleveref}								% Eq., Fig., Table, Section, has to be loaded after hyperref
\creflabelformat{equation}{#2(#1)#3}								% hyperlinks covers the parenthesis around equation
\crefrangelabelformat{equation}{#3(#1)#4 to #5(#2)#6}						% multiple references
\labelcrefformat{subequation}{#2(#1)#3}								% the same for subequations
\labelcrefrangeformat{subequation}{#3(#1)#4 to #5(#2)#6}					% mutiple references

\graphicspath{{./FigureHiperwind_TS}{./FigureHiperwind_SB}}

\begin{document}

\maketitle

\begin{abstract}
	% Probabilistic design of offshore wind turbines should ensure that structural safety is obtained in a cost-effective way, and requires structural reliability assessment of various design options and for different structural responses. 
	The probabilistic design of offshore wind turbines aims to ensure structural safety in a cost-effective way. This involves conducting structural reliability assessments for different design options and considering different structural responses.
	There are several structural reliability methods, and this paper will apply and compare different approaches in some simplified case studies. In particular, the well known environmental contour method will be compared to a more novel approach based on sequential sampling and Gaussian processes regression for an ultimate limit state case study on the maximum flapwise blade root bending moment. For one of the case studies, results will also be compared to results from a brute force simulation approach. Interestingly, the comparison is very different from the two case studies. In one of the cases the environmental contours method agrees well with the sequential sampling method but in the other, results vary considerably. Probably, this can be explained by the violation of some of the assumptions associated with the environmental contour approach, i.e. that the short-term variability of the response is large compared to the long-term variability of the environmental conditions. Results from this simple comparison study suggests that the sequential sampling method can be a robust and computationally effective approach for structural reliability assessment.  
\end{abstract}

\section{Introduction}
Structural reliability assessment is needed to ensure that offshore wind turbines can withstand the environmental conditions they are expected to encounter throughout their lifetimes. Different structural responses are associated with certain limit states that describe under what conditions the structure is expected to fail, and the ultimate limit state (ULS) describes failure when subjected to extreme loads. In probabilistic design, one will typically design against a target reliability level, corresponding to a maximum probability of failure. There are several ways to estimate such small failure probabilities, or conversely, estimate the extreme structural loads and responses associated with a prescribed return period. In this paper, different approaches to structural reliability will be compared by way of a simple offshore wind case study. In particular, the traditional environmental contour approach \cite{WitUCBH93, HW:ENvContLin09} will be compared to a more novel approach based on sequential sampling and Gaussian processes regression \cite{gramstad_sequential_2020}. For one of the case studies, results will be compared to brute-force simulations. The brute force approach is deemed more accurate, but is much more computationally heavy. In fact, the computational burden with this approach made it unfeasible for one of the case studies presented in this paper. 

In this study, the environmental conditions are described by two joint probability distributions of wind speed and turbulence, and a simplified 2-dimensional case study is considered. For one of the case studies, a joint distribution is fitted to wind data for a location near South Brittany, as outlined in \cite{Vanem:OMAE2023, VanemJOMAE24}. In addition, an alternative joint model is assumed for a somewhat different wind climate, without reference to any particular location, but henceforth simply referred to as Site A. In both cases, extreme responses of maximum flapwise blade root bending moment are considered for a floating (South Brittany) and a bottom-fixed (Site A) offshore wind turbine, respectively. In both case studies, a simplified problem is addressed in order to compare the methods, assuming only two environmental input variables; mean wind speed and turbulence at hub height. Hence, the contributions to the dynamic response from wave excitation are not included in the models, which essentially uses a generic onshore wind turbine placed in two offshore environments. 

% brief introduction to the structural reliability problem to be solved and environmental contours. 

Structural reliability analysis is typically performed to determine the reliability ($R$), or conversely the failure probability ($P_f$), of a given design. The limit state function is a performance function that defines when a structure will fail. Generally, the limit state is on the form $g(\boldsymbol{X}) = y_{\text{capacity}} - Y(\boldsymbol{X})$, where $y_{\text{capacity}}$ is the structural strength or capacity, $Y(\boldsymbol{X})$ is the actual load on the structure and $\boldsymbol{X}$ is a vector of relevant input variables. If the loads are greater than the structural capacity, i.e. when $g(\boldsymbol{X}) < 0$, the structure will fail. Hence, the reliability may be determined by integrating the probability density function of the input variables, $\boldsymbol{X}$, over the safe region, i.e. the set of input variables where the structure survives. Denoting the structural performance function $g(\boldsymbol{X})$, and the joint density function of the input variables $f_{\boldsymbol{X}}(\boldsymbol{x})$, the reliability of a structure may be defined as 
\begin{align}
	R = 1 - P_f = P\left[ g(\boldsymbol{X}) > 0 \right] = \int\limits_{g(\boldsymbol{X}) > 0} f_{\boldsymbol{X}}(\boldsymbol{x}) d\boldsymbol{x}.
\end{align}
Such integrals are often difficult to solve exactly since the joint density function of the input (environmental) parameters and the performance function may be complicated functions. Two commonly used methods to approximate these integrals are the First Order Reliability Method (FORM) and the Second Order Reliability Method (SORM), where the failure boundary is approximated by a first- or second order Taylor expansion, respectively, at the design point. 

Often, the environmental input conditions are modelled as piece-wise stationary processes, and the description of the structural response will be a combination of the long-term environmental conditions and conditional short-term responses. Full long-term extreme response analysis would then involve integrating the conditional short-term structural response over all long-term environmental conditions, see e.g. \cite{SND:LongTermResp11, FIG17_LongTermIFORM}. In principle, this may be done by brute force Monte Carlo simulations. However, in practice, the evaluation of the short-term response in a given environmental condition will be very computationally demanding, making this infeasible in most realistic application cases. 

In order to alleviate this, two fundamental approaches can be taken: 1) more efficient short-term response analysis and 2) the need for fewer short-term response calculations. In this study, two approaches for long-term extreme response estimation will be compared, i.e. environmental contours and sequential sampling, which both aim at solving the ULS problem with fewer short term response calculations. However, with both approaches a surrogate model is established for making the short-term response analysis more efficient, but it is noted that this is not strictly necessary and is not an integral part of any of the methods; it is simply applied in this study as a substitute for more computationally demanding calculation methods to speed up short term response calculations. 

When environmental contours are used one essentially solves the \emph{inverse reliability problem}. Rather than computing the failure probability of a given design, one assumes a target failure probability and explores what kind of restrictions this imposes on possible designs; one is essentially separating the description of the environmental input variables from the structural response, see also \cite{WitUCBH93, HW:ENvContLin09}. Hence, environmental contours are contours in the environmental variable space associated with a target exceedance probability of the input variables. 

Under the assumption that the largest response occurs in the most severe environmental conditions, one may define a limit state function as a function of long-term variables only and construct environmental contours in the long-term variable space. One may then assume that any design with a limit state function fully outside of the environmental contour has a failure probability smaller than the exceedance probability associated with that contour. Alternatively, one may assume that the long-term extreme response with a return period corresponding to the exceedance probability of the environmental contour is the maximum response evaluated at points along the environmental contours. Hence, environmental contours can be used to estimate extreme long-term responses for desired return periods, with only a limited set of short-term response analyses. This approach assumes that extreme long-term response is dominated by the long-term variability of environmental condition, and ignores the short-term variability of the response (the conditional variability of the response in an environmental condition), but this may be accounted for by using a higher quantile of the conditional short-term extreme response. A comparison between environmental contours and response-based approaches to reliability assessment of floating structures is presented in \cite{ZD:CompEnvRespSystRel22}. 

Reliability-based design methods for extreme response assessment of offshore wind turbines are proposed in \cite{CvBvKV:RelBaseDesignMethExtremeREsponseOWT03}, where environmental contours are used to identify design load cases, and different statistical methods are used to extrapolate conditional extreme responses. The long-term distribution of extreme response is then the convolution of conditional short-term response distributions over the long-term distribution of sea states. Several other studies have focused on how to best combine the variability of the long-term environmental conditions and the short-term conditional extreme responses to extrapolate and estimate long-term extreme responses of wind turbines, see e.g. \cite{MHB:ExtrEFLoadsProbMeth04, FAM:ImpUndersStatExtrapWTEtreme08, AM:ExtLoadOWTLImitedDAta08, AM:SimOWTRespELP09, SMRPV:StatExtrExtrLoadWT21, YLLZHL:StatExtrapoELWindTurb22}. Datasets for for validation of such techniques have been provided in \cite{Moritary:DBValDLExtrapolTech08}. This is very well aligned with the approaches compared in this paper, which aims at comparing two computationally efficient methods for estimating long-term extreme responses by combining joint distributions of long-term conditions with computationally costly conditional response calculations. This is illustrated by simple case studies, where only the wind conditions are considered for the long-term environmental states and hence ignoring the variability of sea states. For bottom-fixed structures this may be fine; the flap moments may not be notably affected by changes in sea states (as also suggested in \cite{CvBvKV:RelBaseDesignMethExtremeREsponseOWT03}), but it may be an over-simplification for turbines with floating support structures where wave- and wind-induced surge and pitch motions could give rise to inertia forces on the blades (see e.g. \cite{TKN:AeroDynIntWTBladeCFD15, TK:PlatPitchMotFOWTAerodyn15, LH:MotAnalElasticResposFOWT20}). Notwithstanding, the methods for ULS assessment presented in this paper could be extended to also account for long-term variations in sea states, and the simplified case study is still believed to provide a useful relative comparison between the environmental contour and the sequential sampling methods.

It should be noted that the fundamental assumption underpinning the environmental contour method may not be fulfilled for many offshore wind turbine responses, where the structural responses might be non-monotonic due to operational constraints or active and passive structural control \cite{ANS:ControlSystWTLoadStructRel16}. In order to account for this, a modified environmental contour method was proposed in \cite{LGM:ModEnvCont16}, which considers additional contours within different operational regimes, i.e. operational or parked. This approach has later been applied to various offshore wind problems, see e.g. \cite{LGM:ModEnvContSemiSub17, LYGZT:LTExOFFRenEnModEnvCont19, LCOW:ExReFloatVAWTModEnvCont23}, and an extended method was presented in \cite{CJLLR:ExtDnvConOWT20}. A comparison study presented in \cite{LMGM:CompLTExtWindWave18} indicates that the conventional environmental contour method might under-estimate extreme wind-dominated responses compared to the modified contour method. However, in that study, turbulence was not included as an environmental variable, and it was shown in \cite{CJLLR:ExtDnvConOWT20} that including turbulence as a stochastic variable improved results with the original contour method significantly. Indeed, it was found that by including turbulence as an additional stochastic environmental variable the original environmental contour method is able to predict extreme responses from wind-induced loads satisfactorily. In the study presented in this paper, turbulence is indeed included as a stochastic variable, and only the original environmental contour method is included in the comparison study. 

The sequential sampling approach, to be further described in section \ref{sec:SeqSamp} is an alternative approach that aims at minimising the number of short-term response evaluations that accounts for the effect of both long-term and short-term variability on the extreme response. 

%\subsection{Environmental contour method}
The environmental contour method for structural reliability is a well-adopted method, in particular in ocean engineering applications, and it is recommended in DNV's recommended practice on environmental conditions and environmental loads \cite{DNV205} as well as in the NORSOK standard on actions and action effects \cite{NORSOK_N003_17}. Both IFORM (inverse first-order reliability method) \cite{WitUCBH93, HW:ENvContLin09} and DS (direct sampling) \cite{Vanem:EnvCont12, Vanem:EnvCont14} contours are applied to calculate 50-year contours at site A. The output quantity of interest (QoI) considered here is the maximum flapwise blade root bending moment, and the contour method is used to estimate the long-term extreme maximum flapwise blade root bending moment. The different contour methods yield slightly different results, but are generally comparable. Similar results are obtained using a Gaussian process regression and sequential sampling approach \cite{gramstad_sequential_2020}, and also compared with brute force simulation results. The results from the contour methods are generally in line with brute force results if the 90\% fractile of the short-term extreme response distribution is used.  

IFORM contours are applied to obtain extreme response estimates for both 1- and 50-year conditions at the South Brittany site. These are used to estimate the long-term extreme responses of maximum flapwise blade root bending moment for this case study. When compared to the results from the sequential sampling approach, the environmental contour method appears to significantly underestimate the 50-year return value. This is most likely due to the influence of short-term variability of the response, which is found to be quite large for this case study. Brute force calculations were not possible for this case study due to high computational costs. 

\subsection{Problem statement and definitions} 
The goal of this paper is to compare different methods for ULS reliability assessment of offshore wind structures. A simplified case with two environmental input variables is defined, where the response of interest is the maximum flapwise blade root bending moment. The environmental contour method and a sequential sampling method are applied, that may serve as a benchmark to be compared with other approaches in future case studies. It is acknowledged that the 2D case that only considers the variability of the long-term wind conditions is a simplification and the effect of wave induced motions are ignored. Although this might be reasonable for the bottom-fixed structure, it is more questionable for the floating case. Nevertheless, it is believed that the comparison study between the different approaches is still useful, since the same assumptions are made for both approaches. 

The environmental contours approach is a well-established method, that is recommended in DNV's recommended practice \cite{DNV205}. Different variants of this method exist \cite{HW:ENvContLin09, Vanem:EnvCont12, Leira:StocProcCont08, S-GH-ZM-I:EnvContNATAF13, M-IH-Z:EnvContCop15, LGM:ModEnvCont16, LW:DynamicIFORM16, HOWT:HighDensityContour2017, CL:ISORMContours18, MNCCE-GM:AltApproachEnvCont18, ECSADESjointPaper19, HetalBenchmarkREsults21, dHMVanem:QuantCompEC22}, and several applications of this approach are reported in the literature, especially ocean engineering applications \cite{NdLY:EstExtrRespEnvCont98, BM:ApplyContHullLoads01, bhoe:CombContHsT10, MGM:ApplContTwoBodFloatEngConv13, Vanem3Dcontour17, LZD:DesignLoadsFSEnvCont22}, but also for offshore wind turbines \cite{SM:WTExtrLoadIRel04, SM:OWTRelDLEnvCont05, SM:DesiLoadWTEnvCont06, VMA:MultAnalOWT15, VVKS:ProbAnalOffshoreWindTurbExtrRes:EnvCont19, RTGS:UncertAssHydrodynrespoWT20, KHB-P:ExtRespMonoOWTCont24}, earthquake engineering \cite{dLN:EnvContEarthquake00}, bridge design \cite{FIG18_LongTErmPontoon, CFO:EnvContWindBridge22} and ice loads \cite{CLNHE:EnvContIceRidge20}. The Gaussian processes surrogate model with sequential sampling approach is a more recent development \cite{MS:SeqSampExEvent18, gramstad_sequential_2020}, with perhaps less experience although several applications have been reported with such approaches to structural reliability problems as well \cite{EGL:AK-MCS11, AD:SeqBayOptDesignStructRel21, GP:BayExpDesExtREsp22, CFPO:longtermBuffAnalGPMC23, LLW:NewLearnKrigRelEng15, SWLT:LIFKrig17, LBMG:AK-MCSi18, PCGX:AK-SEUR24}; see also the recent overviews in \cite{TNO:AdaptMetaModRelRev21, MMS:ActLearnStrucRel22}. This paper outlines a study where these methods are applied to two offshore wind case studies of ULS reliability assessment, as described in \cite{HIPERWINDWP4.2.1}.   

Response calculations for different environmental conditions are performed with Hipersim (\url{https://gitlab.windenergy.dtu.dk/HiperSim/hipersim}), used to generate turbulence wind boxes assuming the Mann turbulence spectrum, and a fast-to-evaluate surrogate for the NREL 5MW  reference wind turbine \cite{JonkmanNREL} developed by ETH Z\"urich \cite{HiperwindD4.1, schaer2023}, used to calculate the wind turbine responses. It is noted that this surrogate (to be further detailed in section \ref{sec:surrogate} below) is only used as a faster wind turbine simulator and that it is not an integral part of the approaches compared in this study. It is useful to speed up response calculations in this methodological study, but it is noted that the approaches to long-term extreme response assessment would not be dependent on this surrogate. 
% This response calculation is described in more in the subsequent Section~\ref{sec:surrogate}.
% This model is described in \cite{HiperwindD4.1}, and has been trained from OpenFAST \cite{OpenFAST} simulations  of the NREL (National Renewable Energy Laboratory) 5MW  reference wind turbine \cite{JonkmanNREL}. It takes a wind box as input and returns the corresponding maximum flapwise blade root bending moment ($M_y^{\text{Bld}}$).

% \hl{Should we include more details on the surrogates here? E.g. Hipersim and the mNARX models? Possibly ask Styfen and/or Alexis/Nikolay to contribute?}

Two case studies are considered, with offshore wind turbines at two different locations, i.e. Site A and South of Brittany. For both cases, joint statistical models for the relevant input variables were taken as given, but the model for the South Brittany site has been fitted to wind and wave data as outlined in \cite{Vanem:OMAE2023, VanemJOMAE24}. Note, however, that the statistical models were initially fitted to higher-dimensional data, but only mean wind speed and turbulence are considered in this study. 

%\subsection{Teesside}
The joint distribution assumed for site A includes the following parameters:
\begin{itemize}
	\item Mean wind speed ($U$): 10-minute average horizontal wind speed at hub height (83 m).
	\item Turbulence ($\sigma_U$): the temporal standard deviation of the wind speed at hub height.
\end{itemize} 
This is believed to be a realistic description of offshore wind conditions, although it does not refer to a specific location. However, for the purpose of this comparison exercise, it is deemed sufficient. 

%\subsection{South Brittany}
The fitted omnidirectional joint distribution at South Brittany \cite{Vanem:OMAE2023} is considered for the 2-dimensional exercise with the following parameters:
\begin{itemize}
	\item Mean wind speed ($U$): 1-hour average horizontal wind speed at hub height (150 m). This is obtained by down-sampling 10-minute average horizontal wind speed to match the hourly wave conditions for the full joint distribution model. 
	\item Turbulence ($\sigma_U$): the temporal standard deviation of the wind speed at hub height.
\end{itemize} 
This distribution is fitted to hindcast data from the ANEMOC (Digital Atlas of Ocean and Coastal Sea States) database\footnote{URL: http://anemoc.cetmef.developpement-durable.gouv.fr/}, covering 32 years of data from the years 1979 to 2010. It should be noted that these data do not include the turbulence variable, $\sigma_U$, and therefore a conditional distribution for turbulence, conditioned on mean wind speed, was established based on the normal turbulence model described in \cite{IEC-61400-1ed3}, i.e., a conditional lognormal distribution as also outlined in \cite{Vanem:OMAE2023}. For the mean wind speed, a marginal hybrid distribution consisting of a Weibull distribution for the body and a generalized Pareto distribution for the tails was assumed. 

It should be noted that the ETH surrogate model has been trained on the NREL 5 MW turbine whose hub is at 90 m, and the mean wind speed U is not translated to the hub height 90 m in this exercise. Hence, the actual response estimates cannot be used directly, but this is deemed appropriate for a comparison exercise, as long as the same wind input is used in all cases that are to be compared. The response model is only applicable between the cut-in speed (3 m/s) and cut-out speed (25 m/s), so sets of random variables with mean wind speed $U$ below cut-in wind speed (3 m/s) or above cut-out wind speed (25 m/s) were discarded. However, these conditions above the cut-out wind speed are not assumed to contribute to the long-term extreme blade root responses for these turbines, since the turbine will then be in parked conditions, so this is deemed reasonable. 

In the following, the extreme maximum flapwise root bending moment corresponding to a return period of 50 years will be calculated using both environmental contours and a sequential sampling approach for both application cases. The two approaches assume the same type of statistical distribution for the wind variables, and the same models for the response calculations, as outlined above. However, one critical assumption is implicit in the environmental contours approach. That is, it is assumed that the largest responses occur in the most severe environmental conditions. That is, it is tacitly assumed that the effect of the long-term variability of the wind conditions is more important for the extreme response than the short-term variability of the response conditioned on wind conditions. This need not be assumed with the sequential sampling approach, and this study will shed some light on the appropriateness of this assumption for the offshore wind response cases.

\section{Wind turbine response calculation}\label{sec:surrogate}
\subsection{Computational model}\label{sec:comp_model}
The computational model considered in this study is the well-known NREL 5MW onshore wind turbine \cite{JonkmanNREL} equipped with NREL's ROSCO (Reference Open-Source Controller) \cite{ROSCO} and simulated with the open-source aero-servo-elastic simulator code OpenFAST \cite{OpenFAST}. The cut-in, rated and cut-out wind-speeds for this turbine are 3 m/s, 11.4 m/s and 25 m/s, respectively. Further details about this turbine, including natural frequencies and graphs for steady-state responses as a function of wind speed, can be found in \cite{JonkmanNREL}.

Mathematically, we represent this computational model $\mathcal{M}$ through the following equation:
\begin{equation}\label{eq:original_mapping_init_cond}
	y(t) = \mathcal{M}(\bm{x}(\mathcal{T} \le t), \bm{\beta}).
\end{equation}
Here, $\bm{x}$ represents the model input, which is a time-dependent wind field denoted as $\bm{x}(t) \in \mathbb{R}^M$, where $M$ is the dimension of $\bm{x}$ and the time $t$ evolves along a discrete time axis $\mathcal{T} = \{0, \delta t, 2\delta t, \dots, (N-1)\delta t\}$ with a time increment $\delta t$. 
The time series $y(t) \in \mathbb{R}$ is the corresponding model output (e.g. the flapwise blade root bending moment) under the condition that the simulation started at initial conditions $\bm{\beta}$ (e.g. rotor speed, azimuth or initial blade pitch).
Note that we use the notation $\bullet(\mathcal{T} \le t)$ to indicate that the model output can depend on the input up to and including time $t$.

Performing a full aero-servo-elastic (ASE) wind turbine simulation to obtain the turbine response to a given wind input involves substantial computational costs.
Comparing the performance of the environmental contour methods with the sequential sampling approach requires many calls to the ASE simulator, which leads to prohibitively high computational costs. To alleviate the burden associated with these simulations, we replace the expensive simulator with a faster-to-evaluate surrogate model. More detail on the surrogate model is provided in the subsequent Section~\ref{sec:surr_model}.

\subsection{Surrogate model}\label{sec:surr_model}
Surrogate models are often used as fast and accurate approximations for a computationally expensive model $\mathcal{M}$, such as large finite element simulations.
They are typically trained on a small finite set of system inputs $\bm{x}^{(i)}$ and corresponding system outputs $\bm{y}^{(i)}$, also referred to as the \emph{experimental design}, defined as follows:
\begin{equation}\label{eq:experimental_design}
	\mathcal{D} = \left( \bm{x}^{(i)}, \bm{y}^{(i)} \right) , \bm{x}^{(i)} \in \mathbb{R}^{N \times M}, \bm{y}^{(i)} = \mathcal{M}(\bm{x}^{(i)}) \in \mathbb{R}^{N}, i=1, \dots, N_\text{ED}.
\end{equation}
In the context of dynamical systems, such as the aero-servo-elastic simulator, the surrogate $\tilde{\mathcal{M}}$, once trained, shall be able to approximate the output of the system $\mathcal{M}$ at any time instant $t$ at reduced computational cost:
\begin{equation}\label{eq:approx_mappinsg}
	y(t) = \mathcal{M}(\bm{x}(\mathcal{T} \le t), \bm{\beta}) \approx \tilde{\mathcal{M}}(\bm{x}(\mathcal{T} \le t), \tilde{\bm{\beta}}).
\end{equation}
Finding a surrogate model that accurately approximates the response of a dynamical system can be difficult, especially when the system input is high-dimensional or the system response is highly non-linear, which are usual in wind turbine simulations.
In this case, the model input at each time step $t$ is a 2-dimensional spatial plane of wind speed vectors $\bm{x}(t)^{\kappa, \ell} \in \mathbb{R}^3$, where $\kappa, \ell = 1, \dots, 19$ are the lateral and vertical coordinates. 
Furthermore, the turbine has multiple degrees of freedom, some of which are manipulated by the turbine's control systems, resulting in a highly nonlinear input-output mapping.

The surrogate used to replace the simulator in this study is the mNARX surrogate developed in \cite{HiperwindD4.1, schaer2023}.
This is a new class of surrogate models that combine nonlinear autoregressive with exogenous input (NARX) modelling with the sequential construction of an exogenous input manifold. 
While a detailed description of the algorithm is beyond the scope of this paper, the core concepts are presented in the following sections.

\subsubsection{Polynomial NARX model}
The NARX model used in \cite{schaer2023} represents the output at every time step as a sum of monomials $\mathcal{P}_{\bm{\alpha}}$ defined by a set of integer multi-indices $\bm{\alpha}  \in \mathcal{A}$, evaluated on a time-dependent input vector $\bm{\varphi}$ and weighted by a set of real-valued coefficients $c_{\bm{\alpha}}$:
\begin{equation}\label{eq:polynomial_NARX}
	y(t) = \sum_{\bm{\alpha}  \in \mathcal{A}} c_{\bm{\alpha}}  \mathcal{P}_{\bm{\alpha}}(\bm{\varphi}(t)).
\end{equation}
The model captures temporal coherence in the data through the vector $\bm{\varphi}$ containing current and past values of the exogenous inputs $\bm{x}_i(\bullet)$ as well as of past outputs $y(\bullet)$:
\begin{equation}\label{eq:exo_input}
	\begin{split}
		\bm{\varphi}(t) = \{
		&y(t - \ell_{1}^{y}), y(t - \ell_{2}^{y}), \dots, y(t - \ell_{n_y}^{y}), \\
		&x_1(t - \ell_{1}^{x_1}), x_1(t - \ell_{2}^{x_1}), \dots, x_1(t - \ell_{n_{x_1}}^{x_1}), \\ 
		&x_2(t - \ell_{1}^{x_2}), x_2(t - \ell_{2}^{x_2}), \dots, x_2(t - \ell_{n_{x_2}}^{x_2}),    \\
		&\dots, \\
		&x_{M}(t - \ell_{1}^{x_{M}}), x_{M}(t - \ell_{2}^{x_{M}}), \dots, x_{M}(t - \ell_{n_{x_{M}}}^{x_{M}})\},
	\end{split}
\end{equation}
In this formulation $\ell_{i}^{y} \in \{ \delta t, 2\delta t, \dots, (N-1)\delta t \}$ are called autoregressive lags, and $\ell_{i}^{x_j} \in \{0,  \delta t, 2\delta t, \dots, (N-1)\delta t \}$ are the exogenous input lags and they determine the memory of the model. Note that $\ell_{i}^{x_j}$ starts at $0$, opposed to $\ell_{i}^{y}$ which starts at $\delta t$ to preserve causality.
Once the coefficients $c_{\bm{\alpha}}$ of the model are calculated using e.g. least-square minimization as shown in \cite{schaer2023}, the latter iteratively predicts one time-step at a time, similarly to the original simulator:
\begin{equation}\label{eq:NARX_prediction}
	\hat y(t) = \tilde{\mathcal{M}}(\bm{x}(\mathcal{T} \le t), \hat y(\mathcal{T} <t), \tilde{\bm{\beta}}).
\end{equation}

\subsubsection{Manifold construction}
Due to the use of lagged input and output time steps which make the dimensionality of the input vector grow large, NARX models are particularly prone to the curse of dimensionality. As a consequence, in the case of the polynomial NARX models, only low-degree polynomials with few interaction terms can be used limiting their capability to learn complex dynamics.
The mNARX surrogate alleviates this problem by replacing the original input $\bm{x}$ by a manifold $\bm{\zeta}$ to simplify the input-output mapping
\begin{equation}\label{eqn:manifold mapping}
	\tilde{\mathcal{M}}: \bm{\zeta}(\mathcal{T} \le t) \to y(t),
\end{equation}
allowing to better capture nonlinearities, while also allowing to use simpler NARX structures (i.e. low polynomial degree).
It is shown in \cite{schaer2023} that a suitable manifold can be constructed incrementally while incorporating the prior system knowledge usually available when working with physical systems. In this approach they incrementally construct physically meaningfull \emph{auxiliary quantities} $\bm{z_i}$ through the carefully chosen functions $\mathcal{F}_i$:
\begin{equation}\label{eq:aux_quantity}
	\begin{split}
		\bm{z}_{1}(t) &= \mathcal{F}_1(\bm{x}(\mathcal{T} \le t), \bm{z}_{1}(\mathcal{T} <t)) \\
		\bm{z}_{2}(t) &= \mathcal{F}_2(\bm{z}_{1}(\mathcal{T} \le t), \bm{x}(\mathcal{T} \le t), \bm{z}_{2}(\mathcal{T} <t)) \\
		\vdots \\
		\bm{z}_{i}(t) &= \mathcal{F}_i(\bm{z}_{1}(\mathcal{T} \le t), \dots, \bm{z}_{i-1}(\mathcal{T} \le t), \bm{x}(\mathcal{T} \le t), \bm{z}_{i}(\mathcal{T} <t)),
	\end{split}
\end{equation}
eventually forming the exogenous input manifold onto which the final surrogate is constructed:
\begin{equation}
	\hat y(t) = \tilde{\mathcal{M}}^\text{}(\bm{\zeta}(\mathcal{T} \le t),
\end{equation}
with $\bm{\zeta} =\{ \bm{z}_{1}(\mathcal{T}\le t), \dots, \bm{z}_{i}(\mathcal{T} \le t), \bm{x}(\mathcal{T} \le t) \}$. 

Regarding the physically meaningful auxiliary quantities, note that mNARX is a multi-stage surrogate model. It does not directly predict the quantity of interest (QoI) (the blade moment), but first constructs variables that are related to the dynamics of this quantity. These variables can be extracted from the raw input (the wind) or constructed by using another NARX model. They are usually easier to emulate than the main QoI and serve as additional inputs for main NARX model that predicts the QoI. Examples of auxiliary quantities constructed in this study are the blade pitch control system, rotor orientation and rotor speed. For more details of the mNARX surrogate, reference is made to \cite{schaer2023}.

It is demonstrated in \cite{schaer2023} that, thanks to the careful construction of the input manifold, training this mNARX surrogate with a relatively small experimental design comprising 200 points (i.e, 200 ASE simulations) is sufficient to replicate the behavior of the OpenFAST simulator with high accuracy. Furthermore, despite employing a multi-step approach, this emulator is approximately 400 times faster than the original computational model, when used for predicting new output time series.

\section{Extreme responses based on the environmental contour method}
The extreme maximum flapwise root bending moments are estimated by environmental contours for two use cases, as outlined in the following. Essentially, this amounts to establishing environmental contours with a desired exceedance probability and estimating the maximum response for selected environmental conditions along the contours. 

\subsection{Site A}

\begin{figure}[htb]
	\centering
	\includegraphics[width=0.7\textwidth]{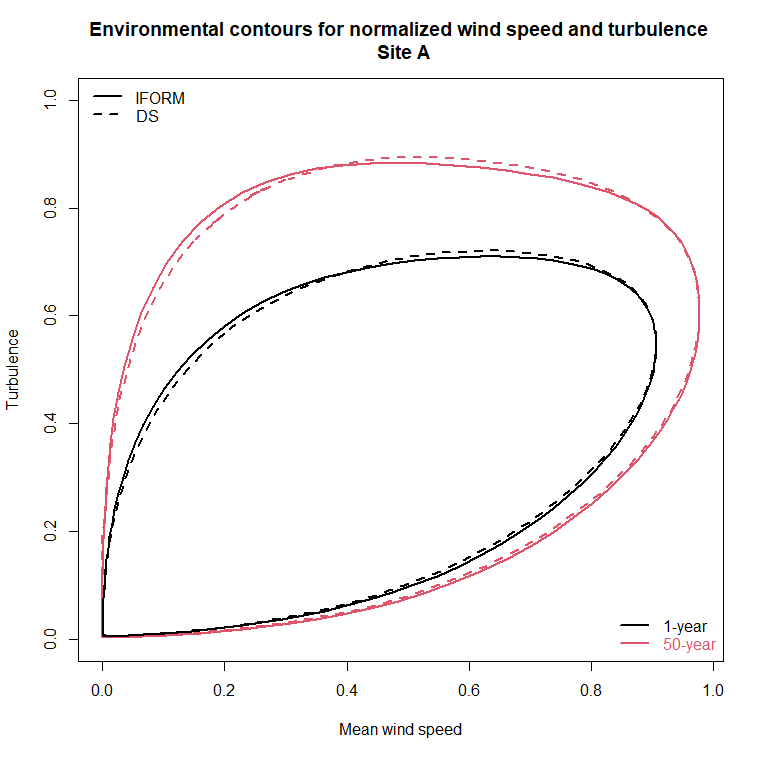}
	\caption{IFORM and DS contours for mean wind speed (wsp ($U$)) and turbulence ($\sigma_U$) based on the assumed distribution for site A.}
	\label{fig:Figure_50yr_ECM}
\end{figure}

To estimate the extreme response of the wind turbine at site A, the 2-dimensional environmental contour method based on DS (direct sampling) \cite{Vanem:EnvCont12, Vanem:EnvCont14} and IFORM (inverse first-order reliability method) \cite{WitUCBH93, HW:ENvContLin09}  is considered. The predefined surrogate model described in \cite{HiperwindD4.1} is used to calculate the short-term extreme response (i.e. maximum flapwise blade root bending moment ($M_y^{\text{Bld}}$)). Environmental contours corresponding to $n$-year extreme of 10-minute conditions are calculated, i.e. corresponding to an exceedance probability of
\begin{equation}
	P_e = \frac{1}{365.25 \times 24 \times 6 \times n}
\end{equation}
For a 50-year return period $n$ = 50 and $P_e$ = 3.8E-07. The DS and IFORM environmental contours based on the assumed joint distribution at site A are shown in Figure \ref{fig:Figure_50yr_ECM} for 1-year and 50-year return periods, respectively. Note that the contours are cropped at the cut-in and cut-out winds speeds of 3 m/s and 25 m/s, since wind conditions outside this range are not believed to contribute to the extreme response. One immediate observation is that the contours from the different contouring methods are quite similar, with only slight differences for both the 1-year and 50-year extreme conditions. 

The points on the 50-year return period contours are taken as input for the surrogate model. There are 68 input points from the 50-year DS-contour, and 40 input points from the 50-year IFORM contour. In total 1000 seeds are run with mNARX (10-minute simulation) for each input point on both the DS-contour and IFORM contour. However, points corresponding to wind conditions below the cut-in wind speed (3 m/s) and above the cut-out wind speed (25 m/s) are disregarded. The response model is not applicable to these conditions, and it is tacitly assumed that these conditions will not contribute to the long-term extreme response. Hence, for wind conditions outside the operational range of the wind turbine, the response is simply set to zero. It is noted that the mNARX model sometimes gives NaN (not a number) for certain seeds for some wind conditions. This is most likely due to negative or extremely high instantaneous winds speeds since this model is not trained neither at negative wind speed nor at extremely high wind speeds. This could occur for example for a combination of high mean wind speed and high turbulence. In such cases, another seed is simply selected to avoid that mNARX gives NaN. This is not believed to significantly impact the final results if this study, and this additional challenge is outweighed by the fact that utilizing this surrogate makes this comparison study feasible. 

The long-term extreme responses of maximum flapwise blade root bending moment (50\% fractile, 90\% fractile, 99\% fractile of the short-term distributions) are taken out from 1000 seeds based on 50-year DS and IFORM contours, respectively. That is, for each wind condition along the contours, N = 1000 response simulations are performed, and the distribution of maximum responses from these 1000 simulations are used to extract the quantiles of interest. According to \cite{DNV205}, which quantile to use is highly case-dependent and there are no clear recommendations of what quantile to consider. Hence, Results for three different quantiles are reported. The results are listed in Table \ref{tab:ExtremeResponse}.  It appears that the long-term extreme estimations of flapwise blade root bending moment are comparable for the two contour methods when considering the 50\% fractile, i.e., 20.59 MNm using the DS contours and 20.40 MNm from the IFORM contours, with a similar combination of mean wind speed ($U$) and turbulence ($\sigma_U$). The results are also close for the 90\% fractile based on these two approaches, i.e., 25.85 MNm from the DS approach and 25.17 MNm from the IFORM with a similar combination of mean wind speed and turbulence. As for the results from 99\% fractile, even though the extreme responses do not differ too much, i.e., 34.01 MNm from the DS approach and 34.96 MNm from IFORM, the combination of mean wind speed and turbulence is quite different, i.e., the mean wind speed leading to the extreme response according to the DS approach is 22.74 m/s  and the corresponding turbulence is 5.17 m/s  while the mean wind speed leading to the extreme response for IFORM is 13.14 m/s  and the corresponding turbulence is 5.30 m/s. According to DNV-RP-C205 \cite{DNV205}, extreme values in a random process are random variables with a statistical variability. Generally the relevant factor and fractile will be larger for strongly nonlinear problems, and an appropriate high fractile should be chosen for the characteristic long-term response. The appropriate fractile level is case-specific, and a fractile in the order of 85\% to 95\% will often be a reasonable choice for use in design. In this exercise, the results from the contour method will later be compared with results from brute force simulations in order to consider which quantile levels are most appropriate. 

%==============================
\begin{table}[htp]
	\centering
	\caption{Long-term extreme responses of the maximum flapwise blade root bending moment $M_y^{\text{Bld}}$ [MNm] with estimated 50-year return period based on DS and IFORM, respectively; Site A.}
	\label{tab:ExtremeResponse}
	\centering
	\begin{tabular}{c c c c c c c c c}
		\hline
		\hline
		\multicolumn{3}{c}{Direct sampling} \\
		\hline
		$U$ [m/s] & $\sigma_U$ [m/s]  & $M_y^{\text{Bld}}$ [MNm] \\
		\hline
		12.66  & 5.33  & 20.59 (50\% fractile)\\
		14.56  & 5.36  & 25.85 (90\% fractile)\\
		22.74  & 5.17  & 34.01 (99\% fractile)\\
		\hline
		\multicolumn{3}{c}{IFORM} \\
		\hline
		$U$ [m/s] & $\sigma_U$ [m/s] & $M_y^{\text{Bld}}$ [MNm] \\
		\hline
		13.14  & 5.30  & 20.40 (50\% fractile)\\
		14.51  & 5.30  & 25.17 (90\% fractile)\\
		13.14  & 5.30  & 34.96 (99\% fractile)\\
		\hline
		\hline
	\end{tabular}
\end{table}

\begin{figure}[htb]
	\centering
	\includegraphics[width=1.0\textwidth]{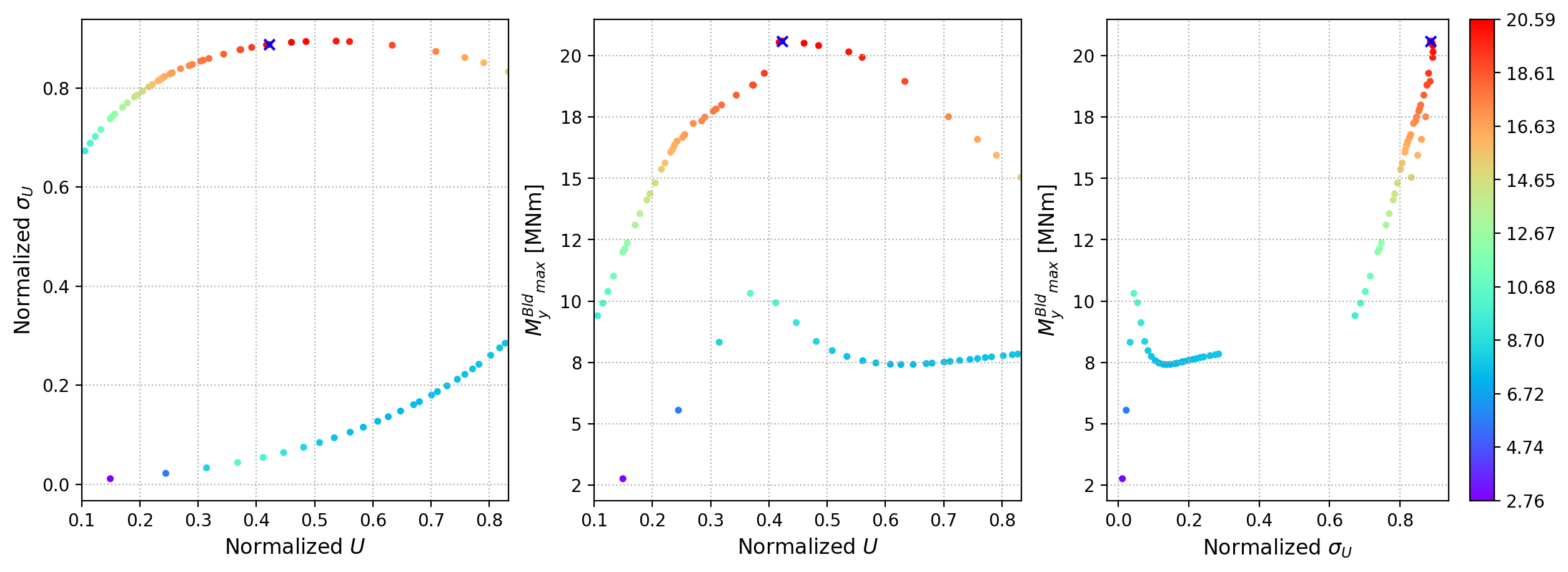}
	\caption{DS 50-year contour and corresponding extreme response 50\% fractile of flapwise blade root bending moment $M_y^{\text{Bld}}$ [MNm]; Site A.}
	\label{fig:Figure_50yr_DS_P50}
\end{figure}

\begin{figure}[htb]
	\centering
	\includegraphics[width=1.0\textwidth]{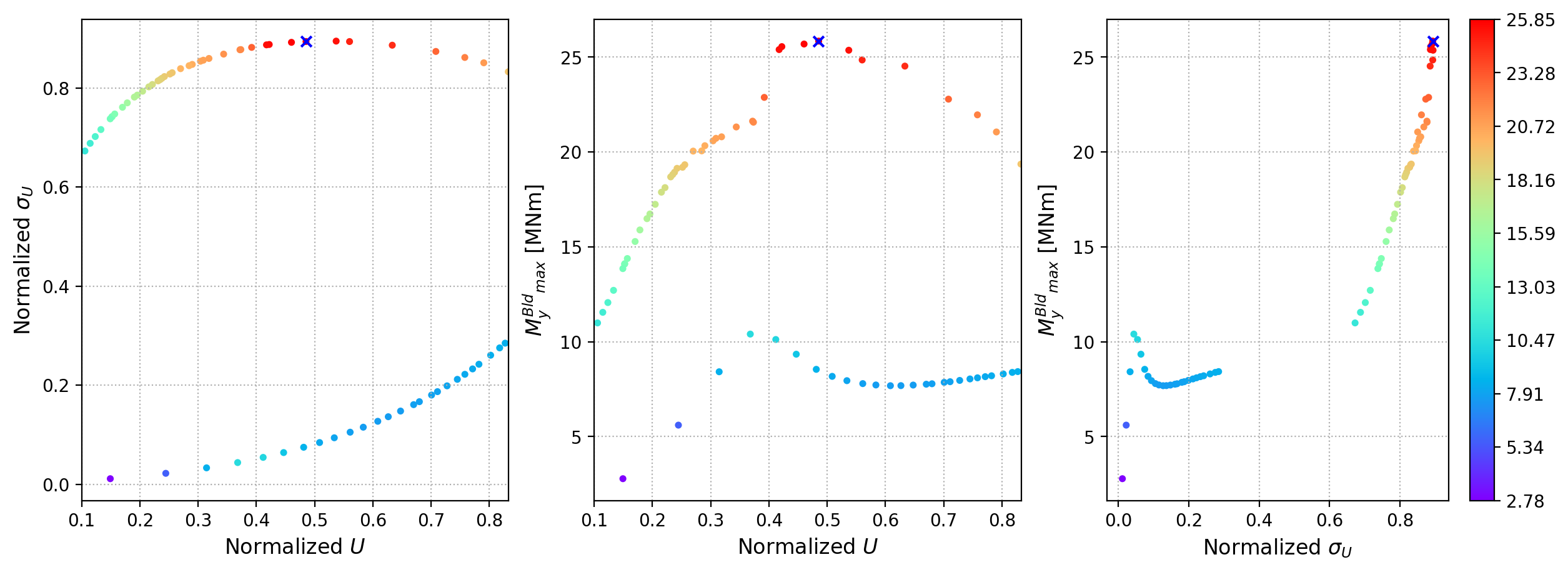}
	\caption{DS 50-year contour and corresponding extreme response 90\% fractile of flapwise blade root bending moment $M_y^{\text{Bld}}$ [MNm]; Site A.}
	\label{fig:Figure_50yr_DS_P90}
\end{figure}

\begin{figure}[htb]
	\centering
	\includegraphics[width=1.0\textwidth]{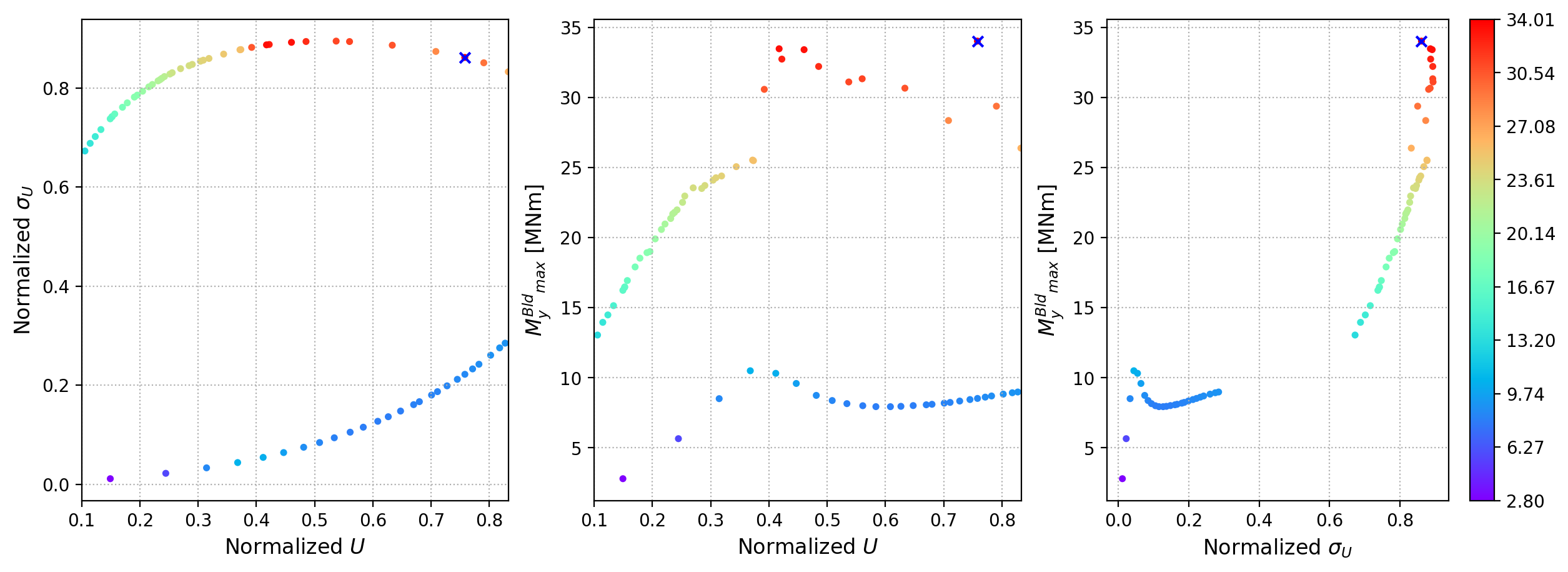}
	\caption{DS 50-year contour and corresponding extreme response 99\% fractile of flapwise blade root bending moment $M_y^{\text{Bld}}$ [MNm]; Site A.}
	\label{fig:Figure_50yr_DS_P99}
\end{figure}

Figures \ref{fig:Figure_50yr_DS_P50} to \ref{fig:Figure_50yr_DS_P99} show the 50-year DS-contour and the corresponding long-term extreme response of the maximum flapwise blade root bending moment by using the 50\% fractile, 90\% fractile and 99\% fractile of the short-term extreme response distributions, respectively. The rainbow color denotes the range of the maximum value and the blue cross denotes the combination of the wind speed and turbulence leading to the maximum extreme flapwise blade root bending moment along the 50-year DS-contour.

\begin{figure}[htb]
	\centering
	\includegraphics[width=1.0\textwidth]{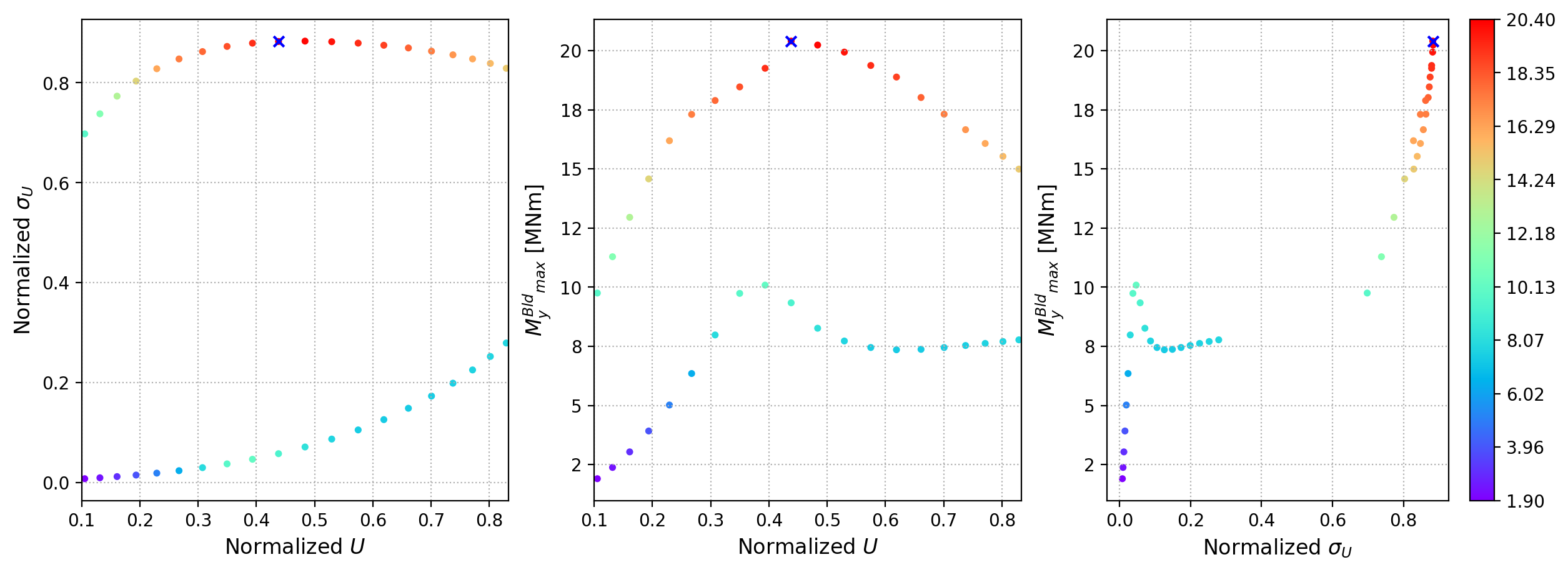}
	\caption{IFORM 50-year contour and corresponding extreme response 50\% fractile of flapwise blade root bending moment $M_y^{\text{Bld}}$ [MNm]; Site A.}
	\label{fig:Figure_50yr_IFORM_P50}
\end{figure}

\begin{figure}[htb]
	\centering
	\includegraphics[width=1.0\textwidth]{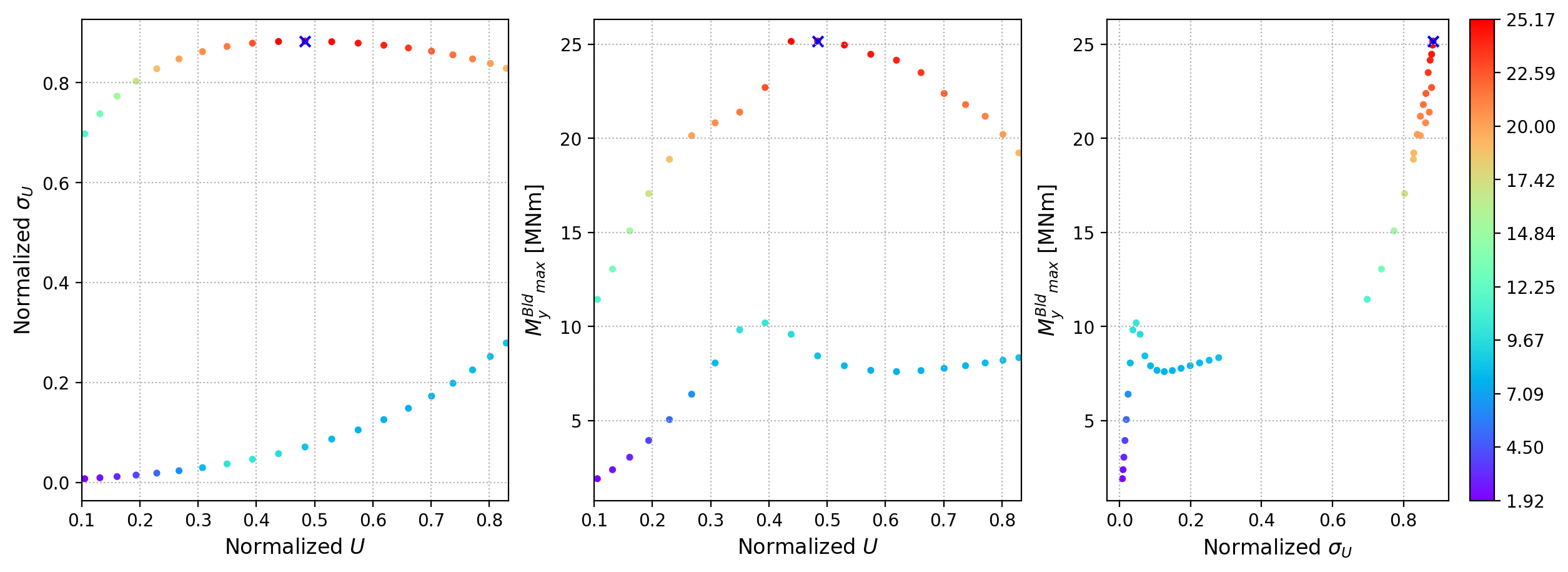}
	\caption{IFORM 50-year contour and corresponding extreme response 90\% fractile of flapwise blade root bending moment $M_y^{\text{Bld}}$ [MNm]; Site A.}
	\label{fig:Figure_50yr_IFORM_P90}
\end{figure}

\begin{figure}[htb]
	\centering
	\includegraphics[width=1.0\textwidth]{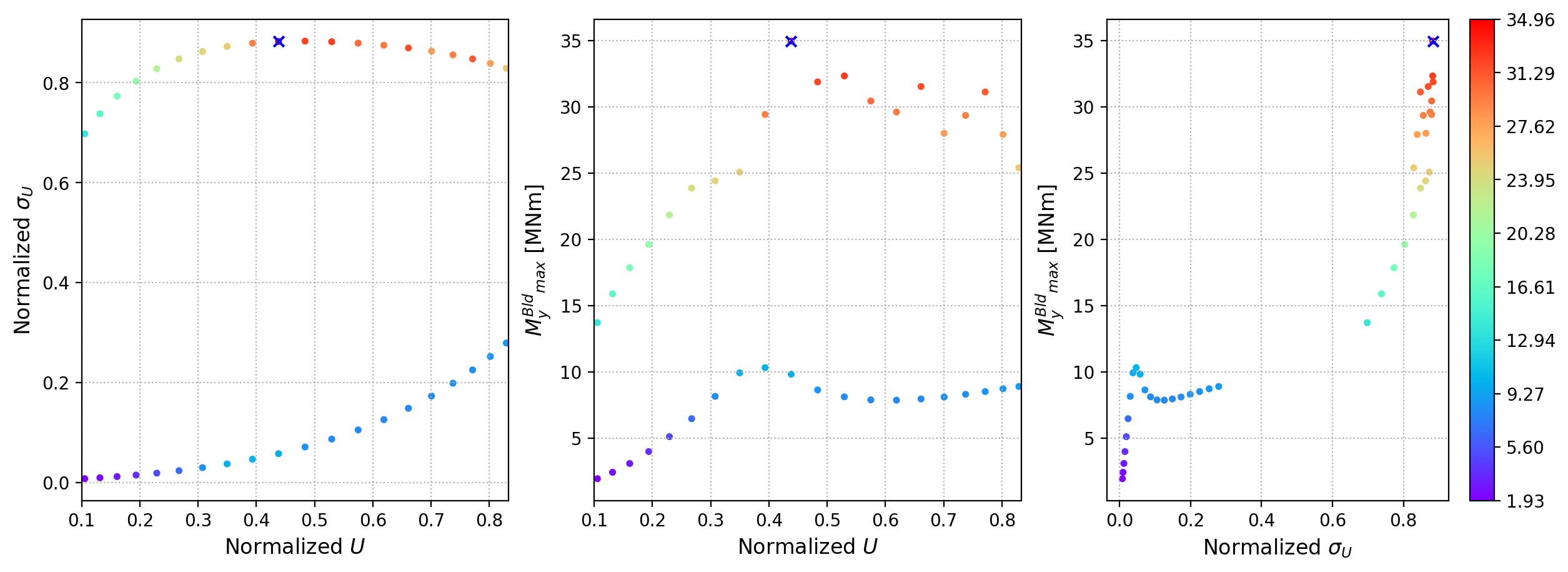}
	\caption{IFORM 50-year contour and corresponding extreme response 99\% fractile of flapwise blade root bending moment $M_y^{\text{Bld}}$ [MNm]; Site A.}
	\label{fig:Figure_50yr_IFORM_P99}
\end{figure}

Figures \ref{fig:Figure_50yr_IFORM_P50} to \ref{fig:Figure_50yr_IFORM_P99} show the 50-year IFORM contour and the corresponding estimated long-term extreme response of the maximum flapwise blade root bending moment by using 50\% fractile, 90\% fractile and 99\% fractile, respectively, of the short-term maximum response distribution. The rainbow color denotes the range of the maximum value, and the blue cross denotes the combination of the wind speed and turbulence leading to the maximum flapwise blade root bending moment on the 50-year IFORM contour.

It is observed that for both contour methods, the response tends to be largest for relatively moderate mean wind speeds around  12 - 15 m/s, but that it tends to grow with turbulence and gets its largest values for the highest values of turbulence. 

%==============================
%==============================

%==============================
%==============================
%==============================

\subsection{South Brittany}
\begin{figure}[htb]
	\centering
	\includegraphics[width=0.7\textwidth]{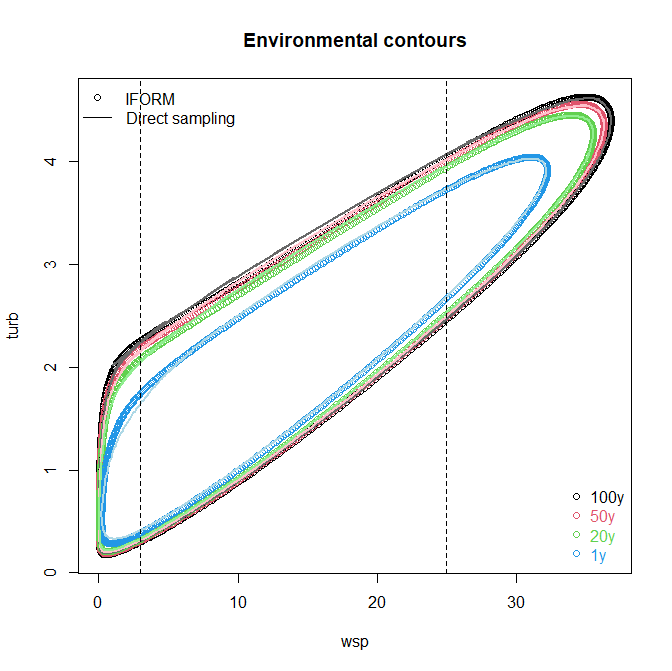}
	\caption{IFORM and DS contours for wsp ($U$) and turbulence ($\sigma_U$); South Brittany location}
	\label{fig:EC_South_Brittany}
\end{figure}

To estimate the extreme response of the wind turbine at South Brittany, 2-dimensional environmental contours based on IFORM (inverse first-order reliability method) are considered for the mean wind speed and turbulence variables. The predefined surrogate model described in \cite{HiperwindD4.1} is used to calculate the short-term extreme response (i.e. maximum flapwise blade root bending moment ($M_y^{\text{Bld}}$)) in selected wind conditions. Environmental contours corresponding to $n$-year extreme of 1-hour conditions are calculated, i.e. corresponding to an exceedance probability of
\begin{equation}
	P_e = \frac{1}{365.25 \times 24 \times n}
\end{equation}
For 1- and 50-year return period ($n$ = 1, 50), $P_e$ = 1.14E-04 and 2.28E-06, respectively. The IFORM environmental contours based on the fitted omnidirectional joint distribution at South Brittany are shown in Figure \ref{fig:EC_South_Brittany} for 1- and 50-year return periods. The points along the 1- and 50-year IFORM contours are taken as input for the surrogate model. There are 94 input points from the 1-year IFORM contour, and 74 input points from the 50-year IFORM contour. In total 100 seeds are run with the surrogate (1-hour simulation) for each input point on the IFORM contours. The long-term extreme responses of maximum flapwise blade root bending moment (50\%, 90\% and 99\% fractiles) are taken out from 100 seeds based on 1- and 50-year IFORM contours, respectively. The results are listed in Table \ref{tab:ExtremeResponseSB}. Note that the 99\% fractile should be used with caution since these results are based on only 100 seeds and the 99\% fractile estimate is therefore not very reliable. Notwithstanding, the results are included in the table, and the estimates corresponding to the 90\% fractile are presented in bold font, since this seems to be the most reasonable choice. 

%==============================
%==============================

%==============================
\begin{table}[htp]
	\centering
	\caption{Long-term extreme responses of flapwise blade root bending moment $M_y^{\text{Bld}}$ [MNm] based on 1- and 50-year IFORM contours, respectively. South Brittany case.}
	%    \caption{Table: South Brittany.}    
	\label{tab:ExtremeResponseSB}
	\centering
	\begin{tabular}{c c c c c c c c c}
		\hline
		\hline
		\multicolumn{3}{c}{IFORM - 1year } \\
		\hline
		$U$ [m/s] & $\sigma_U$ [m/s] & $M_y^{\text{Bld}}$ [MNm] (1-h) \\
		\hline
		11.11 &	2.57 & 14.95 (50\% fractile) \\
		\textbf{11.11} & \textbf{2.57} & \textbf{15.87 (90\% fractile)} \\
		10.63 & 2.53 & 16.88 (99\% fractile) \\
		\hline
		\multicolumn{3}{c}{IFORM - 50 year} \\
		\hline
		$U$ [m/s] & $\sigma_U$ [m/s] & $M_y^{\text{Bld}}$ [MNm] (1-h) \\
		\hline
		13.31 & 3.04 & 15.53 (50\% fractile) \\
		\textbf{ 12.69} &	\textbf{2.99} & \textbf{16.54 (90\% fractile)} \\
		13.31 & 3.04 & 17.54 (99\% fractile) \\
		\hline
		\hline
	\end{tabular}
\end{table}
%==============================

\begin{figure}[htb]
	\centering
	\includegraphics[width=1.0\textwidth]{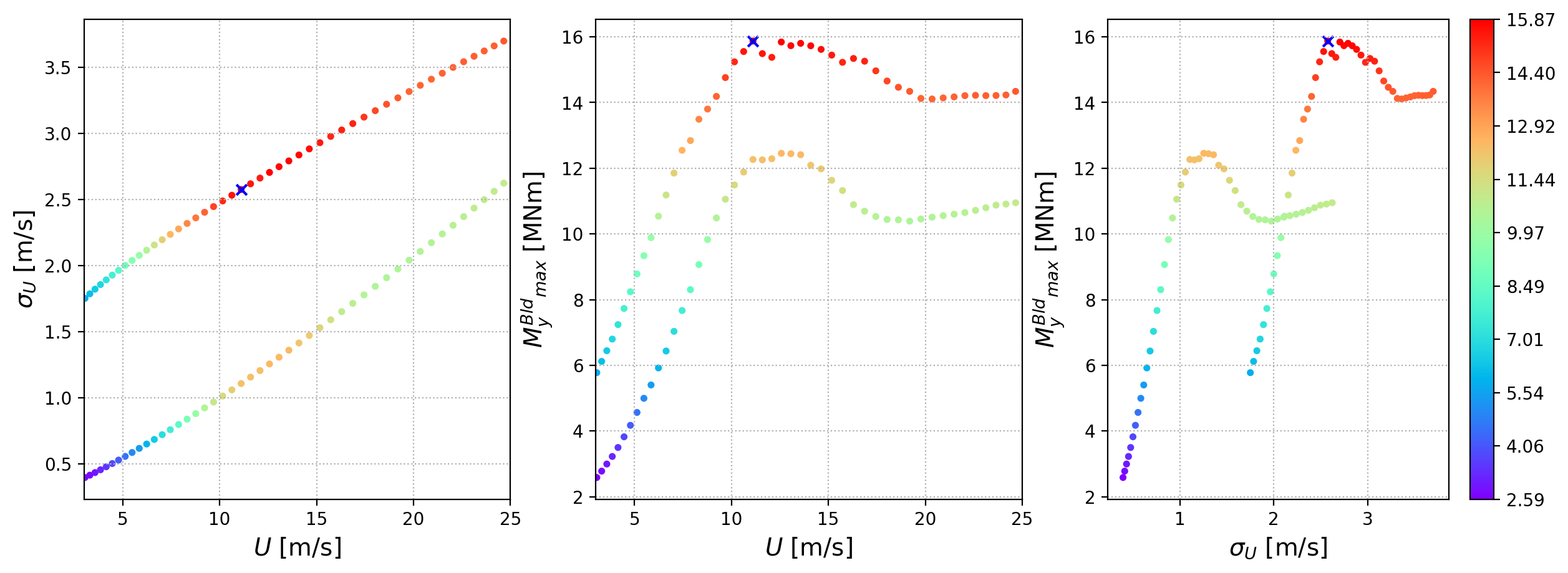}
	\caption{IFORM 1-year contour and corresponding extreme response 90\% fractile of flapwise blade root bending moment $M_y^{\text{Bld}}$ [MNm]; South Brittany case.}
	\label{fig:IFORM_UsdU_1y_P90}
\end{figure}

\begin{figure}[htb]
	\centering
	\includegraphics[width=1.0\textwidth]{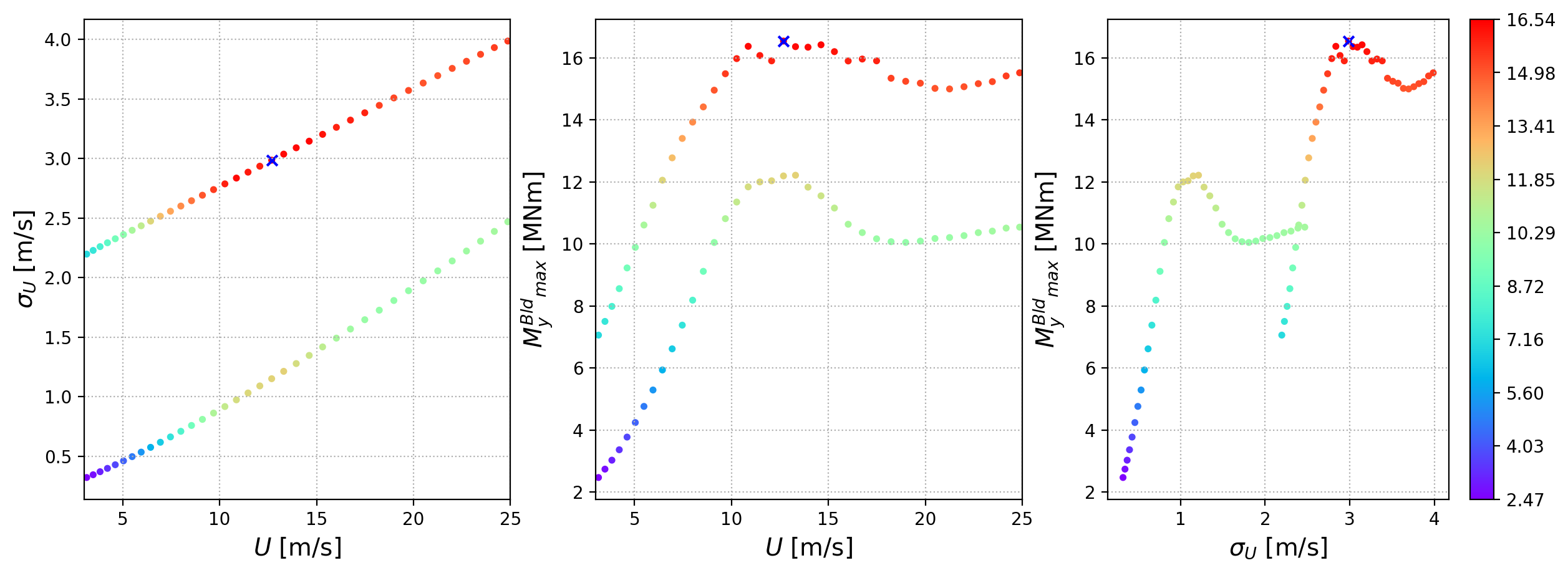}
	\caption{IFORM 50-year contour and corresponding extreme response 90\% fractile of flapwise blade root bending moment $M_y^{\text{Bld}}$ [MNm]; South Brittany case.}
	\label{fig:IFORM_UsdU_50y_P90}
\end{figure}

The long-term extreme estimations of flapwise blade root bending moment from the 50-year contour is slightly higher than the long-term extreme estimations of flapwise blade root bending moment from the 1-year contour, assuming the 90\% fractile, i.e., 15.87 MNm from 1-year contour and 16.54 MNm from 50-year contour with a similar combination of mean wind speed and turbulence. Figures \ref{fig:IFORM_UsdU_1y_P90} and \ref{fig:IFORM_UsdU_50y_P90}  show the 1- and 50-year IFORM contour and the corresponding long-term extreme response of the maximum flapwise blade root bending moment by using 90\% fractile, respectively. The rainbow color denotes the range of the maximum value, and the blue cross denotes the combination of the wind speed and turbulence leading to the maximum flapwise blade root bending moment on the IFORM contour.

%==============================
%==============================

\section{Estimating extreme structural response using Gaussian Process regression and sequential sampling} \label{sec:SeqSamp}
\subsection{Problem description}
In the following we consider the problem of estimating the extreme structural response of a wind turbine under wind-loading using Gaussian process regression and sequential sampling. The problem is governed by the long-term stochastic variability of the environment and the short-term stochastic variability of the structural response. See \cite{HIPERWINDWP4.2.1} for a more detailed problem description.

The long-term environment is here described in terms of $\xb=(U, \sigma_{U})$, where $U$ is the average horizontal wind speed at hub height and $\sigma_{U}$is the turbulence intensity defined as the temporal standard deviation of the wind speed. The long-term parameters $\xb$ are assumed stationary over a period of 10 minutes for the Site A case and for 1 hour for South Brittany, and are described by a joint probability distribution $f_{\Xb}(\xb)$, as presented in \cite{HiperwindD2.3}.

The short-term response of interest is the maximum flapwise blade root bending moment and is for a given long-term condition $\xb$ a stochastic process $S_t(t)$ for $t\in[0, 10]$ min. In the following we will consider the maximum response during each 10-minute time-period, in the following denoted $Y=\max_{t\in[0, 10]}S_t(t)$. Hence, for a given long-term input, $Y$ is a stochastic random variable from an (unknown) probability distribution $g_{Y|\Xb}(y|\xb)$.

The marginal long-term distribution of the maximum blade root bending moment can then in principle be found by integrating over the long-term environment, i.e.
\begin{align}\label{eq:lt_integral}
	g_{Y}(y) = \int g_{Y|\Xb}(y|\xb)f_{\Xb}(\xb)d\xb.
\end{align}
In the present case the distribution $g_{Y|\Xb}(y|\xb)$ is not known, but samples from $g_{Y|\Xb}(y|\xb)$ can be obtained by running simulations of the structural response, which in this case is provided by an mNARX surrogate model. In principle, the integral \eqref{eq:lt_integral} can then be estimated through a brute-force Monte-Carlo approach. However, this is still a computational demanding task, and may in many cases be impossible if simulating the structural response is computational demanding.

\subsection{Method description}
In the following we apply the methodology described in \cite{gramstad_sequential_2020}, see also \cite{MS:SeqSampExEvent18} to estimate the integral \eqref{eq:lt_integral}. The main steps in the methodology can be summarized as follows:
\begin{enumerate}
	\item Introduce a parametric distribution $\hat{g}_{Y|\Xb}(y|\xb, \thb(\xb))$ for the short-term response (i.e. $\hat{g}_{Y|\Xb}(y|\thb(\xb))$ represents an approximation of the "true" distribution $g_{Y|\Xb}(y|\xb)$). 
	\item A Gaussian process (GP) regression model is used to represent the distribution parameters $\thb(\xb)$, which are fitted based on a limited number of short-term response simulations.
	\item The estimated long-term response distribution is obtained from \eqref{eq:lt_integral} by replacing $g_{Y|\Xb}(y|\xb)$ with $\hat{g}_{Y|\Xb}(y|\thb(\xb))$.
\end{enumerate}
More details are given in the following subsections.
\subsubsection{Long-term model}\label{sec:lt_model}
The long-term models used for the present case studies are the joint distribution of the the mean wind speed $U$ and turbulence intensity $\sigma_{U}$ described in \cite{HiperwindD2.3, Vanem:OMAE2023}. In evaluation of the integral \eqref{eq:lt_integral}, a period of $N_y=10 000$ years are considered, sufficient to provide reliable estimates of the 100- and 50-year return values of interest. Hence, $N=N_y\cdot 365.25\cdot 24 \cdot 6$ samples from the long-term simulations are used.

For the present problem it has been assumed that conditions with mean wind speeds $U<3$ m/s and $U>25$ m/s do not contribute to the extreme response. Hence, these regions of the long-term distributions has been ignored in the analysis, equivalent to assuming that these wind speeds give zero response. For the high wind speed this is justified by the fact that the turbine is shut-down for large wind speeds above 25 m/s.

\subsubsection{Parametric model for the short-term response}\label{eq:st_reponse}
In the present work two different models $\hat{g}_{Y|\Xb}(y|\thb(\xb))$ for the short-term extreme response have been considered: the Gumbel-distribution and the generalized extreme value (GEV) distribution. The Gumbel distribution has two parameters $\thb=(\alpha, \beta)$ (location and scale) and the GEV distribution has three parameters $\thb=(\alpha, \beta, \gamma)$ (location, scale and shape). Note that this corresponds to the MAX approach according to \cite{CvBvKV:RelBaseDesignMethExtremeREsponseOWT03}, where only maxima are used to extrapolate the conditional short-term response distributions; see also \cite{RM:StatExtrETExtLoad08, TSV:AssLoadExtrapolWT11, Vanem:OEME15} for a discussion on different methods for extrapolating short-term conditional extreme responses.  

From $n_{seeds}$ random response simulations for a given long-term input $\xb$, the best fit distribution parameters $\theta(\xb)$ are found as the maximum likelihood estimate (MLE) for the given observations $\bm{y}=(y_1, \cdots, y_{n_{seeds}})$. An important part of the present methodology is to include the uncertainty in the distribution parameters throughout the analysis. This is achieved by considering the likelihood of the distribution parameters under the given observations, i.e. $p(\bm{y}|\thb)$. In order to incorporate this uncertainty into the GP-model, the Gumbel and GEV likelihoods are approximated by 2- and 3-dimensional Gaussian likelihoods, respectively.

\begin{figure}[htb]
	\centering
	\includegraphics[width=1.0\textwidth]{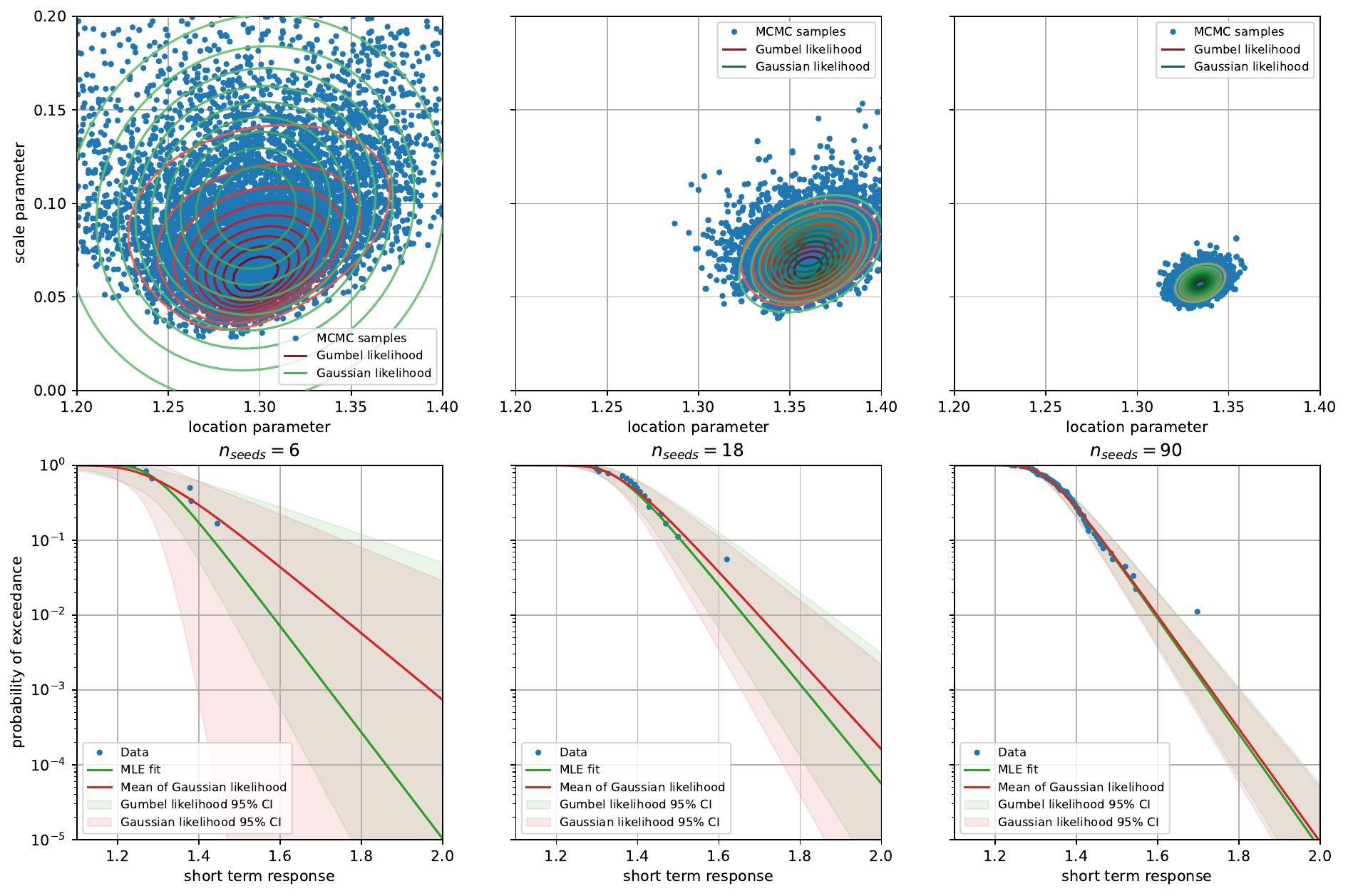}
	\caption{Examples of Gaussian likelihood fitting to account for uncertainty in parameter estimates. Upper row: The "true" likelihood of the Gumbel-parameters under the observed data, the corresponding MCMC-samples, and the resulting best-fit Gaussian likelihoods. Second row: The corresponding Gumbel-distributions with 95\% confidence intervals based on the "true" and fitted likelihoods.}
	\label{fig:likelihood_fit}
\end{figure}

The best-fit Gaussian likelihood is found by drawing samples from the distribution proportional to $p(\bm{y}|\thb)$ using Markov Chain Monte Carlo (MCMC). From the MCMC-samples, the means and covariance matrix of the distribution parameters are estimated. The set of MCMC samples is increased in batches until three consecutive estimates of the means and covariances are within 1\% of each other. This procedure is illustrated in Fig.~\ref{fig:likelihood_fit}, for different numbers of data ($n_{seeds}$ short-term simulations) used to fit the distributions. The upper row of Fig.~\ref{fig:likelihood_fit} shows the "true" likelihood of the Gumbel-parameters under the observed data, the corresponding MCMC-samples, and the resulting best-fit Gaussian likelihoods. The second row shows the corresponding Gumbel-distributions, with 95\% confidence intervals for the "true" and fitted Gaussian likelihoods. As expected, the uncertainty in the distribution parameters decreases with more data. So does the consistency of the Gaussian fit, which becomes much better for larger $n_{seeds}$.

\subsubsection{GP-model for distribution parameters}
The fitted Gaussian likelihood for the distribution parameters (i.e. the mean vector and covariance matrix) is then used to fit a GP model. In the following, we consider the general case that the Gaussian process have $m$-dimensional output (i.e. models $m$ distribution parameters $\thb=(\theta^{(1)}, \cdots, \theta^{(m)})\in\mathbb{R}^m$) and $d$-dimensional input (i.e. is a function of $d$ long-term parameters $\xb\in\mathbb{R}^d$). In the present case $m=2$ or $m=3$ for the Gumbel and GEV distributions, respectively, and $d=2$ for $\xb=(U, \sigma_U)$.

That is, we consider a Gaussian process given as a prior over functions $\thb: \mathbb{R}^d\to \mathbb{R}^m$:
\begin{align}\label{eq:gp_prior}
	\thb(\xb)\sim GP(\mub(\xb), K(\xb, \xb')),
\end{align}
where the prior mean $\mub(\xb)=[\mu_{1}(\xb),\cdots, \mu_{m}(\xb)]\tran$ is here assumed zero, and where $K$ is the diagonal matrix
%\begin{align}
%	K(\xb, \xb')=	
%	\begin{bmatrix}
%		K_{1}(\xb, \xb') & 0 & \cdots & 0 \\
%		0 & K_{2}(\xb, \xb') & \cdots & 0 \\
%		\vdots & \vdots & \ddots & \vdots \\ 
%		0 & 0 & \cdots &  K_{m}(\xb, \xb')
%	\end{bmatrix},
%\end{align}
\begin{align}
	K(\xb, \xb')=	
	\begin{bmatrix}
		K_{1}(\xb, \xb') & 0  & 0 \\
		0 & \ddots &  0 \\
		0 & 0 &  K_{m}(\xb, \xb')
	\end{bmatrix},
\end{align}
where each $K_j$ here is of the Mat\'ern 3/2 type as in \cite{gramstad_sequential_2020}.

Given some training data, i.e. observed distribution parameters $\zetab_j=(\zeta_j^{(1)}, \cdots, \zeta_j^{(m)})$ for points $\xb_j$: $D=\{(\xb_j, \zetab_j)\}_{j=1}^{N}$ one can derive the posterior predictive distribution for "new" points under the observed training data. As described in section \ref{eq:st_reponse} the distribution parameters are assumed to come with Gaussian noise, so that $\zetab_j=\theta(\xb_j)+N(0, \Sigma_j)$, where $\Sigma_j$ is the covariance matrix of each set of the $m$ distribution parameters, as estimated using the procedure described in section \ref{eq:st_reponse}. For further details on the actual implementation it is referred to the Python source code \cite{agrell_2023}.

\subsubsection{Simulation of response from GP-model}
Given the GP-model that enables drawing random samples of the distribution parameters for any long-term parameter $\xb$, a "full" Monte-Carlo estimate of the response distribution is obtained. First, distribution parameters $\thb_j$ are sampled from the GP-model for each long-term condition $x_j$, $j=1,\cdots, 10000\cdot 365.25\cdot 24 \cdot 6$ in the 10 000-year long-term simulation. Then, for each $\thb_j$ a short-term response $y_j$ is sampled from the Gumbel- or GEV distribution $\hat{g}_{Y|\Xb}(y|\thb_j)$. From the 10 000 years of responses, the relevant return values are estimated. 

\subsubsection{Sequential update of the GP-model}
\begin{figure}[htb]
	\centering
	\includegraphics[width=1.0\textwidth]{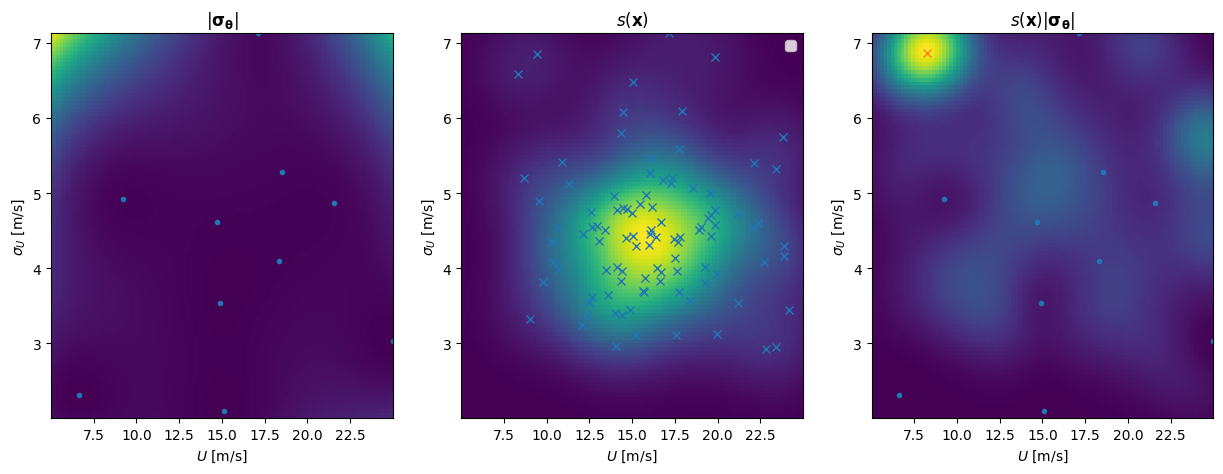}
	\caption{Left: The standard deviation of distribution parameters as function of long-term parameters $\xb=(U, \sigma_{U})$. Middle: The estimated distribution $s(\xb)$ of responses above 100-year return value). Right: The acquisition function $s(\xb)|\boldsymbol{\sigma}_ {\thb}(\xb)|$. Blue dots show existing points used to train the GP-model, blue crosses show points leading to a response above the 100-year return value, and the red cross shows the new point $\xb_{new}$.}
	\label{fig:new_point_method}
\end{figure}

As described in \cite{gramstad_sequential_2020}, a sequential update of the GP-model is applied, where a new point $\xb_{new}$ for which to run new short-term response simulations is selected based on a trade-off between increasing accuracy in the areas of the long-term input space that contributes to the extreme response (here responses above the estimated 100-year level) and areas where the uncertainty is large. More specifically, the following acquisition function is applied
\begin{align}\label{fig:acq_fun}
	\xb_{new}=\argmax_{\xb} s(\xb)|\boldsymbol{\sigma}_ {\thb}(\xb)|,
\end{align}
where $\boldsymbol{\sigma}_{\thb}(\xb)=(\sigma_1(\xb),\cdots, \sigma_m(\xb))$ are the standard deviations of each of the distribution parameters as function of the long-term variables $\xb$, and $s(\xb)$ is the probability density function of responses above the 100-year return value, which is estimated using kernel density estimation. This approach is illustrated in Fig.~\ref{fig:new_point_method}, which shows the acquisition function \eqref{fig:acq_fun} used to select the new point to add to the GP-model.

In the following, results from applying this method to estimate the long-term extreme maximum flapwise blade root bending moment for offshore wind turbines at two locations -- Site A and South Brittany, respectively, will be presented. 

\subsection{Results Site A}
In the following, results from the sequential sampling approach are presented in terms of the 50- and 100-year return values, which in the following are estimated as the $1-1/50$ and $1-1/100$ fractiles of the distribution of the annual maximum response, respectively. The empirical distribution of the annual maximum response is, as explained above, estimated from a 10 000-year simulation in each iteration of the sequential sampling approach.

In addition, the failure probability $p_f$, defined as the annual probability that the response exceeds 27.112 MNm is considered \cite{HIPERWINDWP4.2.1}.

\subsubsection{Brute-force results}
\begin{figure}[htb]
	\centering
	\includegraphics[width=1.0\textwidth]{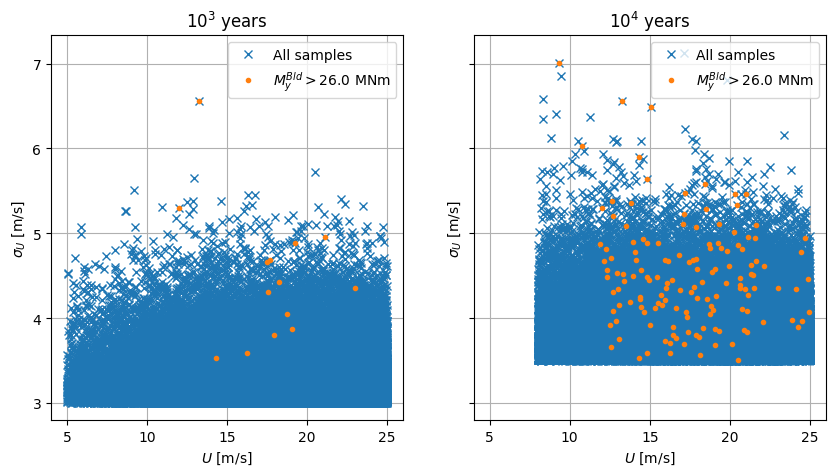}
	\caption{Brute force samples for the $10^3$-year and $10^4$-year periods.}
	\label{fig:brute_force_samples}
\end{figure}

In order to have a reference to which to compare the sequential sampling results, brute-force estimations of the 100-year return value for $M_y^{Bld}$ were carried out, through a direct Monte-Carlo approach. While the mNARX surrogate model is too computationally expensive to carry out an intensive sample of the full 10 000-year period, a reliable estimate could be obtained by excluding long-term conditions from which there are none or very small contributions to the 100-year return value. Two sets of brute-force samples were carried out, applying two different truncations of the long-term distribution, as summarized in Tab.~\ref{tab:brute_force}. Effectively, the truncation of the long-term distribution can be viewed as a simple form of importance sampling where it is implicitly assumed that responses outside the cutoff regions are zero. Hence, only return levels for which there are minor contribution from the discarded long-term conditions could be reliably estimated. However, as seen from Fig.~\ref{fig:brute_force_samples}, this is a valid assumption for $M_y^{Bld}$ larger than about 26.0. Hence, it is believed that the 100-year return value of 27.112, as obtained from the 10 000-year of samples is a reasonable accurate estimate. The estimate based on 1 000 year of data is obviously even less affected by the truncation of the long-term-distribution, but more affected by statistical uncertainty.
\begin{table}[htb]	
	\centering	
	\caption{Overview of the brute-force estimation of the 100-year return value for the response $M_y^{Bld}$.}
	\label{tab:brute_force}		
	\begin{tabular}{ c c c c c}
		duration & cutoff $\sigma_U$  & cutoff $U$  &  100-year $M_y^{Bld}$ & 50-year $M_y^{Bld}$ \\\relax
		[years] & [m/s] & [m/s] &  [MNm] & [MNm] \\\hline
		1 000 & 3.0 & 5.0 & 26.334 & 24.649\\  
		10 000 & 3.5 & 8.0 & 27.112 & 25.093 \\\hline
	\end{tabular}
\end{table}

\subsubsection{Sequential sampling results}
\begin{figure}[htb]
	\centering
	\includegraphics[width=1.0\textwidth]{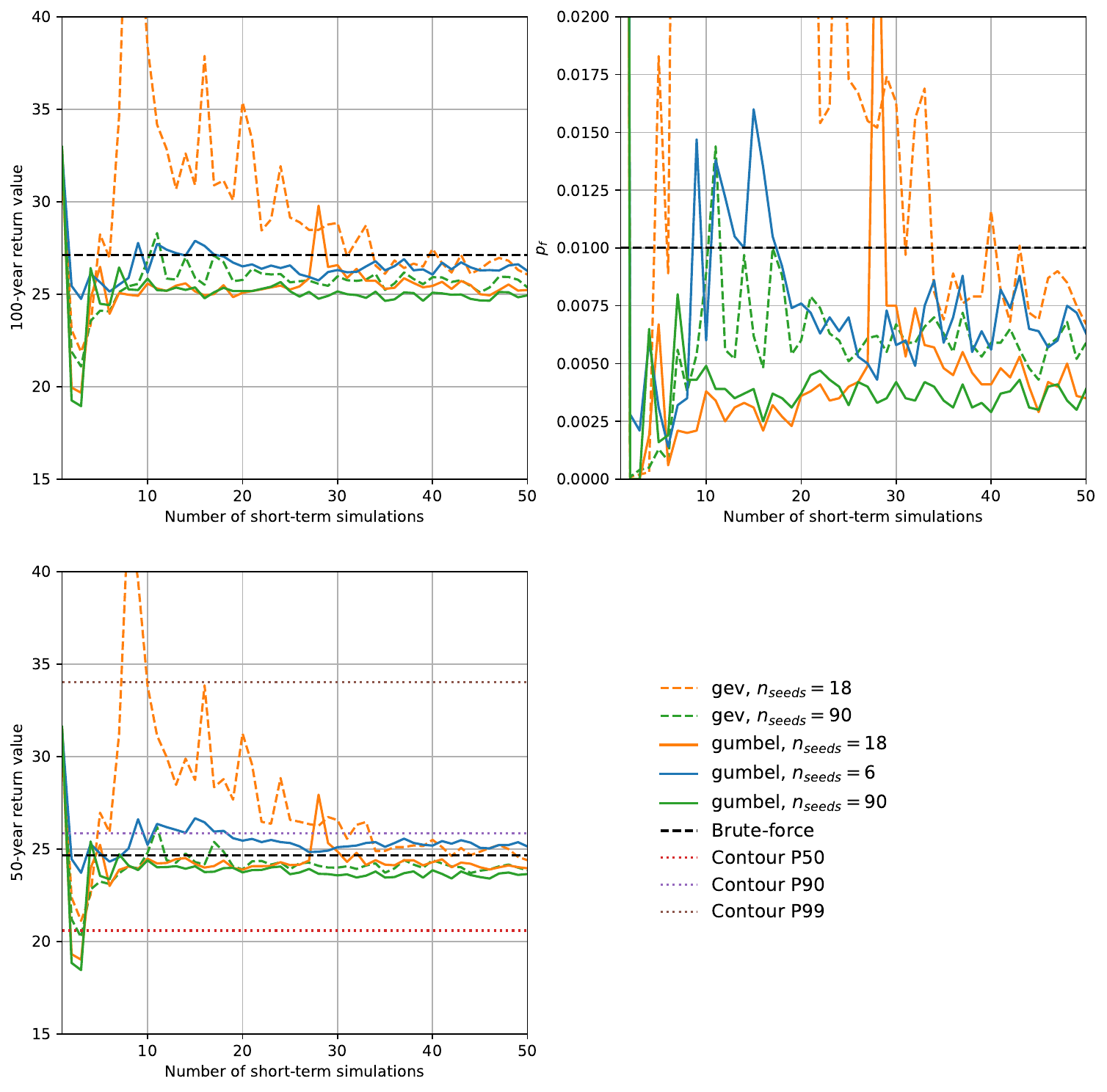}
	\caption{Estimated 100- and 50-year return values for Site A as well as estimated failure probability $p_f$ from sequential sampling, as function of the number of short-term simulation used to train the GP. Results from brute-force sampling and from the direct sampling contour method (for 50-year RV) are also included.}
	\label{fig:seq_samp_res_TS}
\end{figure}

As described in section \ref{eq:st_reponse}, the GP-model represents the distribution parameters in the distribution of the 10-minute maximum response. The distribution parameters are fitted based on $n_{seeds}$ 10-minute simulations. Different distribution models and different number of seeds (i.e. number of 10 minute response simulation) were tested. Below results for the Gumbel distribution for 6, 18 and 90 seeds (i.e. 1-hour, 3-hours and 15-hours of response simulations in each long-term condition) and the GEV distribution for 18 and 90 seeds are reported. The results, in terms of estimated 100- and 50-year return values as well as failure probability $p_f$, as function of the number of short-term simulation (i.e. the number of long-term conditions used to train the GP) are shown in Fig.~\ref{fig:seq_samp_res_TS}. 
As seen from Fig.~\ref{fig:seq_samp_res_TS} the estimated 50- and 100-year return values converges to values in relatively good agreement with the brute-force results within 10-20 iteration of the sequential sampling approach. For the case using the GEV-distribution with $n_{seeds}=18$, the convergence is somewhat slower, which can be explained by the fact that large uncertainties in the distribution parameter estimates (i.e. large variation when sampling distribution parameters from the GP leads to large variations of the responses). 

Compared with the environmental contour estimates, it is observed that assuming the 90\% quantile of the short-term response distribution, contour estimates agrees reasonably well with the brute-force and the sequential sampling approach. Assuming the 99\%-quantile significantly overestimates the 50-year extreme response, whereas assuming the median response leads to a significant underestimation of the extreme response. However, the 90\%-quantile agrees well, with only a slight overestimation of the response, which corresponds to slightly conservative designs. 

\subsection{Results South Brittany}
\begin{figure}[htb]
	\centering
	\includegraphics[width=1.0\textwidth]{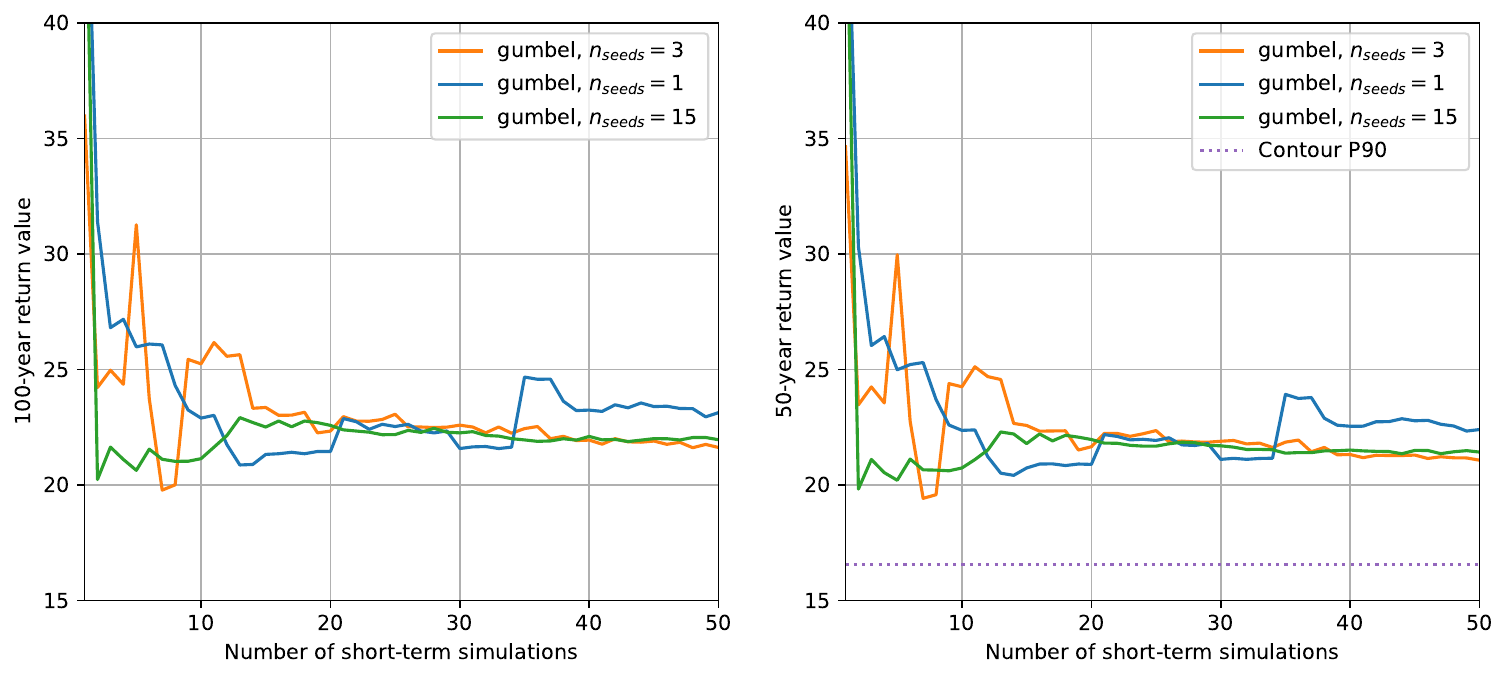}
	\caption{Estimated 100- and 50-year return values for South Brittany from sequential sampling, as function of the number of short-term simulation used to train the GP. For the 50-year return period the result from the IFORM contour method is shown.}
	\label{fig:seq_samp_res_SB}
\end{figure}

For the South Brittany case the results of the sequential sampling are shown in Fig.~\ref{fig:seq_samp_res_SB}. Here, only the Gumbel distribution was considered. Note also that for South Brittany the period over which the long-term variables $(U, \sigma_U)$ are assumed stationary is one hour, so that e.g. $n_{seed}=3$ in Fig.~\ref{fig:seq_samp_res_SB} corresponds to running three one-hour response simulations. The Gumbel distribution was, however, still fitted to 10-minutes maximum responses as for Site A. That is, each one-hour response time series was split in six 10-minutes parts, from which the 10-minute maximum values were extracted.

\begin{figure}[htb]
	\centering
	\includegraphics[width=0.45\textwidth]{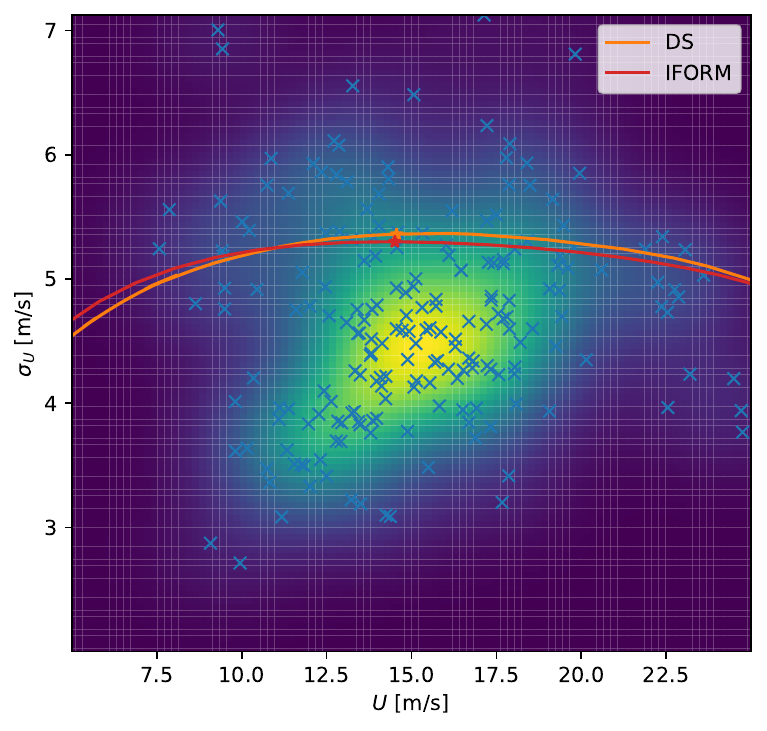}
	\includegraphics[width=0.45\textwidth]{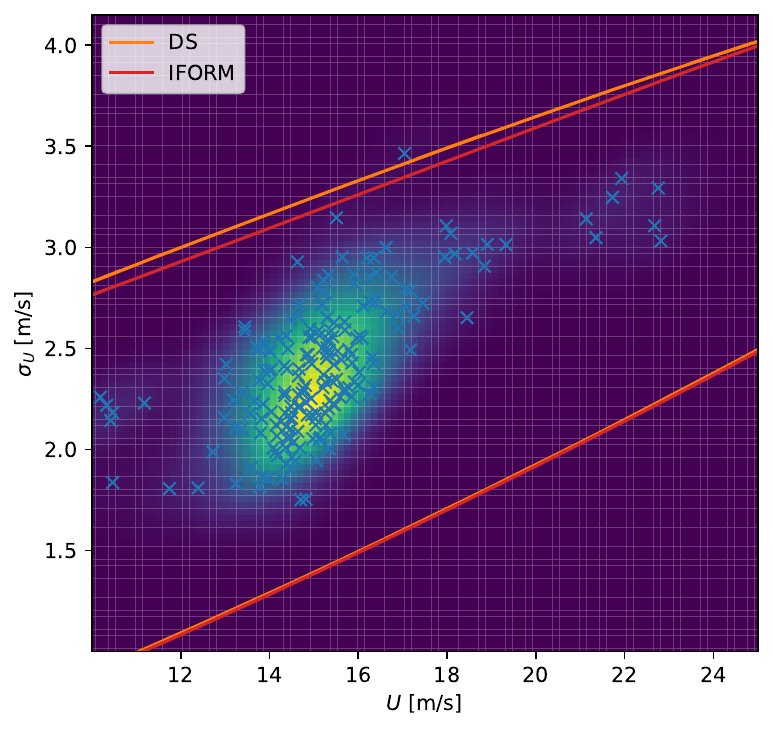}
	\caption{Contribution to the 50-year return value as function of long-term parameters for Site A (left) and South Brittany (right).}
	\label{fig:50yr cont}
\end{figure}

It is interesting to note that for South Brittany the contour approach underestimates the 50-year return value significantly. It is likely that this due to the fact that the main contribution to the 50-year response is coming from long-term parameters well inside the contour. Hence, this represents a situation where extreme short-term responses in relatively common long-term conditions dominates the extreme response. This is illustrated in Fig.~\ref{fig:50yr cont}, which shows the area in the long-term plane that has responses exceeding the 50-year return level. It is observed that for Site A, the long-term conditions giving rise to extreme responses are much closer to the contour lines compared to South Brittany. It is not clear exactly why the short-term variability is so much more dominating in the South Brittany case, but one possible explanation could be that in this case study, hourly stationary wind conditions were assumed, whereas 10-minutes stationary conditions were assumed in the Site A case.

\clearpage
\section{Summary and conclusions}
In this study, different approaches to ULS reliability assessment have been applied to two offshore wind cases and compared. In both cases, the extreme maximum flapwise blade root bending moment has been estimated by environmental contours and by a Gaussian processes regression model with a sequential sampling approach. These results may be used as a reference and for comparison with other approaches to ULS of the same problem. 

For one of the use cases -- the Site A case -- it is observed that the contour method, provided a suitable quantile of the short-term extreme response distribution is selected, is in reasonable agreement with the sequential sampling approach, and also with results from a brute force simulation exercise. That is, assuming the 90\% quantile of the short-term extreme response distribution, the contour-based estimates of the 50-year extreme response is about 25 MNm (25.85 MNm for a direct sampling contour method and 25.17 MNm for an IFORM-based contour). This is in reasonable agreement with the brute force results of the same response (24.65 MNm from 1000 years of simulations and 25.09 MNm from 10 000 years of simulations). Also, the sequential sampling approach yields estimates within this range, and in most cases the sequential sampling approach converges quickly to a reasonable estimate; typically within 10 - 20 iterations in most cases. Overall, for this use case, the environmental contours method appears to slightly overestimate the 50-year extreme response, whereas the sequential sampling approach sometimes slightly overestimate and sometimes slightly underestimates the response, assuming that the brute force estimate is close to the true extreme response. 

The Site A case study applied contours from two different contour-methods, i.e., direct sampling and IFORM contours. For this case study, results from the different contour methods are in reasonable agreement with only minor differences. However, this is also case-dependent, and previous studies have demonstrated that there might considerable differences between contour methods in other cases \cite{VanemB-G:JOMAE15, VanemContourCompStruc17, ECSADESShipCase19}, see also \cite{ECSADESjointPaper19, HetalBenchmarkREsults21}. 

This use case also illustrates the effect of the short-term response. With both contour-methods, estimates assuming different quantiles of the short-term maximum response distribution were obtained, and these range between around 20 MNm (from the median short-term maximum response) to almost 35 MNm (from the 99\% quantile of the short-term maximum response distribution). This difference is significant, and illustrates the importance of selecting an appropriate quantile for the short-term extreme response when applying the contour method. As this is case-specific, it may be difficult to select this a priori, and calibration against a full long-term analysis may be needed \cite{DNV205}. Notwithstanding, for this particular case study, it appears that the 90\% quantile level is appropriate and yields reasonable results. 

The sequential sampling approach, on the other hand, directly accounts for the short-term variability in the response, and does not need to determine an appropriate quantile level. This is deemed a huge advantage of this approach, especially in cases where the short-term variability is notable.  

The motivation for both the environmental contour method and the sequential sampling approach is that short-term response calculations are computationally heavy and time consuming; it may not be feasible to perform a full long-term analysis. Hence, it is relevant to compare the two approaches in terms of computational efficiency. This may be compared by considering how many combinations of the long-term wind parameters are needed as input for the different approaches. With the brute force simulation approach, some cutoffs were applied in order to reduce the number of required calculations, but still a very large number of short-term response calculations were needed. With the environmental contour methods, between 40 and 68 points along the contours were used. However, realizing that extreme responses are not expected for some parts of the contours (i.e. the lower parts corresponding to relatively small turbulence), these numbers could easily be reduced by at least half. Nevertheless, the environmental contour methods still require short-term response calculations for in the order of 20 - 30 long-term wind conditions. Results for the same case using the sequential sampling approach indicate that results converge within the same number of short-term simulations. Hence, in terms of computational efficiency, the two approaches are comparable. 

In summary, for the Site A case study, the contour method and the sequential sampling methods are found to be comparable both in terms of accuracy and computational efficiency. However, the fact that the sequential sampling method accounts for the short-term variability of the response, without the need to determine an appropriate quantile level of the short-term distribution provides a strong argument for preferring this method. 

The second case study -- South Brittany -- only included one contour method (IFORM), and also did not include brute force results since this was too computationally costly (essentially, because the cut-off values in the long-term parameter space would be too small). However, this case study illustrates potential difficulties and shortcomings with the environmental contour method when the short-term variability is large. Results from the environmental contours and the sequential sampling do not concur; the contour estimates of the 50-year response are significantly lower than the estimates from the sequential sampling approach. Most likely, this is because the short-term variability is dominating in this case study, violating the implicit assumptions of the environmental contour method. This is confirmed by Fig.~\ref{fig:50yr cont}, which shows that for the South Brittany case, the wind conditions most likely to be responsible the long-term extreme response lie well within the environmental contour lines. They correspond to recurrent operational conditions that will not be identified by contour methods. Hence, the assumption that the largest response will occur in the most severe wind conditions is not true. Indeed, due to large short-term variability, significant contributions to the long-term extreme response come from frequently occurring non-extreme wind conditions. This observation is also substantiated by the observation that there are relatively small differences between the 1-year and the 50-year extreme response estimates. 

In this second case study, the true response is not known since brute force simulations were not feasible, but given the importance of the short-term variability, it is safe to assume that the sequential sampling results are closer to the true values. Hence, in this case study, the environmental contour approach fails to deliver reliable results, whereas the sequential sampling results are deemed more reasonable. One remedy for the environmental contour approach is to increase the quantile level of the short-term response distribution, but the exact level to choose can be difficult to determine. Alternatively, one may inflate the contours \cite{DNV205}, but again, it may be difficult to determine an appropriate inflation factor. Hence, also this case study suggests that the sequential sampling approach should be favoured. However, without knowledge of the true extreme response, final recommendations cannot be made with high confidence. 

One possible explanation for the relatively stronger influence of the short-term variability of the response in one of the case studies (South Brittany) can be that for this case study, a lower number of seeds was used for the response calculations for each environmental condition. If the extreme loading is dominated by extrapolation of the loads, a lower number of seeds could yield larger variability of the extremes. Hence, using 100 seeds rather than 1000 seeds is a possible explanation for the difference in short-term variability. Notwithstanding, even 100 seeds is a relatively large number and it is believed to correspond to a sufficient number short term response calculations. Further investigation into the effect of number of seeds is recommended in actual design applications. 

Further case studies for other structural problems are recommended in order to validate the sequential sampling approach to ULS reliability assessment in general. It will be of interest to investigate how this method performs for higher-dimensional problems where the structural response depends on more environmental variables. This will be the focus of further work, where the simplified 2-dimensional case study will be extended to higher-dimensional cases.

%----------------------------------------------------------------------------------------
\section{Acknowledgements} 		

This work has been carried out within the HIPERWIND project funded by the European Union’s Horizon 2020 Research and Innovation Programme under Grant Agreement No. 101006689. %We would like to thank  Styfen Sch{\"a}r and Alexis Cousin for providing the mNARX surrogate model which was used to significantly speedup aero-servo-elastic (ASE) simulations, and therefore enable fully quantitative wind turbine design practices.

\bibliography{bibliography.bib}

\end{document}